\documentclass{emulateapj}
\usepackage{amssymb,amsmath,amsthm}

\newcommand{\xmm}{\textit{XMM-Newton}}
\newcommand{\chandra}{\textit{Chandra}}
\newcommand{\rosat}{\textit{ROSAT}}
\newcommand{\suzaku}{\it Suzaku\rm}
\newcommand{\euve}{\it EUVE\rm}

\newcommand{\pspc}{\textit{PSPC}}

\def\as1101{Abell~S1101}

\shorttitle{Soft excess in AS1101}
\shortauthors{Bonamente et al.}

\begin{document}


\title{X-ray spectroscopy of Abell~S1101 with Chandra, XMM-Newton and ROSAT: 
band-pass dependence of the temperature profile and 
soft excess emission}


\author{Massimiliano Bonamente\altaffilmark{1,2} and Jukka Nevalainen\altaffilmark{3,4} }

\altaffiltext{1}{Physics Department, University of Alabama in Huntsville,
Huntsville, Al 35899}
\altaffiltext{2}{NASA National Space and Technology Center, Huntsville, Al 35899}
\altaffiltext{3}{Finnish Center for Astronomy with ESO, University of Turku, 
V\"ais\"al\"antie 20, FI-21500, Piikki\"o, Finland}
\altaffiltext{4}{Department of Physics, Dynamicum, P.O. Box 48, 00014 University of Helsinki, Finland}


\begin{abstract}
We present spatially-resolved spectroscopy of the galaxy
cluster \as1101, also known as S\`{e}rsic 159-03,  with
\chandra, \xmm\ and \rosat,  and investigate the presence of
soft X-ray excess emission above the contribution from the hot intra-cluster medium.
In earlier papers  we reported an extremely bright soft excess component
that reached 100\% of the thermal radiation in the R2 \rosat\ band (0.2-0.4 keV), using
the HI column density measurement by Dickey and Lockman. In this paper
we use the newer Leiden-Argentine-Bonn survey measurements
of the HI column density towards \as1101,  significantly lower than the
previous value, and show that the soft excess 
emission in \as1101\ is now at the level of 10-20\% of the hot gas emission, in line
with those of a large sample of clusters analyzed by Bonamente et al. in 2002.  
The \rosat\ soft excess emission is detected  regardless
of calibration uncertainties between \chandra\ and \xmm.
This new analysis of \as1101 indicate that the 1/4~keV band emission 
is compatible with the presence of WHIM filaments
connected to the cluster and extending outward into the intergalactic medium; the
temperatures we find in this study are typically lower than the WHIM probed in other X-ray studies.
We also show that the soft excess emission is 
compatible with a non-thermal origin as the inverse Compton scattering
of relativistic electrons off the cosmic microwave background, with pressure
less than 1\% of the thermal electrons.
\end{abstract}




\section{Introduction}
 Evidence for an excess of soft X-ray radiation in clusters above the contribution from the virial gas 
 was initially discovered by \citet{lieu1996a,lieu1996b} in the
extreme ultra-violet ($h \nu \sim 0.1$ keV) with \euve, and then confirmed with several other instruments,
notably in the \rosat\ 1/4 keV band (see  \citealt{bonamente2002} for the analysis
of a sample of 38 clusters, and \citealt{durret2008} for a recent review of the literature).
The excess emission is usually modelled as an additional thermal component of lower temperature
($kT \sim 10^6-10^7$~K) or as a non-thermal power law. 
A plausible scenario for the soft excess 
is thermal emission from  warm filaments
seen in projection towards clusters \citep[e.g.,][]{mittaz2004,bonamente2005}, or
from low-entropy dense gas within the cluster \citep{cheng2005}.
Alternatively, a non-thermal power law can be used to model the excess,
 indicative of relativistic electrons that scatter the
CMB photons and emit by Compton scattering \citep[e.g.,][]{sarazin1998,lieu1999}.
Other explanations are also viable, e.g. multi-temperature structure due to 
several mergers during the lifetime of a cluster \citep{lehto2010}.
Given the limited spectral resolution of the current CCD 
detector technology, it has not been possible to prove conclusively which additional model
is a better fit to the data \citep[see, e.g.,][]{nevalainen2003,bonamente2005,werner2007}.

Among the factors that affect the detection of the soft excess emission,
the column density of absorbing HI plays a primary role. In recent years, the Leiden-Bonn-Argentine
survey (LAB) of \cite{kalberla2005} has effectively superseded the \cite{dickey1990} measurements,
and significant differences between the two surveys are occasionally present. 
Among the clusters in which the
excess has been reported in the literature, \as1101\ is an outstanding case in that the
\cite{dickey1990} value of $N_H=1.80 \times 10^{20}$ cm$^{-2}$ differs significantly from the
LAB value of $N_H=1.15 \times 10^{20}$ cm$^{-2}$. A lower $N_H$ value implies less Galactic absorption, and therefore
more of the detected soft X-ray flux from the cluster originates as the soft tail of the hot, virial
gas, reducing the need for an additional emission component. 

The goal of this paper is to perform a spatially-resolved 
spectroscopic study of \as1101\ using 
the available data from \chandra\ and \xmm, needed to measure the temperature of the hot gas, and
\rosat, needed to test whether at soft X-ray energies the radiation is consistent with the low-energy tail of the
hot gas, or whether an excess of emission is present. This paper improves on the initial
detection of the \rosat\ soft excess in \as1101\ 
by \cite{bonamente2001a} in two ways: the hot gas is measured from \chandra\ and \xmm\ data that are much more sensitive 
to hot gas than the \rosat\ observations alone, and the use of the new LAB value for $N_H$.
The presence of a soft component in the spectrum is investigated by testing whether 
the best-fit temperature of the hot gas has a bandpass dependence, as observed in certain clusters
\citep[for example,][ see Section~\ref{sec:tprofiles}]{cavagnolo2008,lehto2010}, by  modelling the hot gas
with a single-temperature model and a two-temperature model to account for
cooler gas seen in projection, and determine whether the soft X-ray flux lies above the
prediction (Section~\ref{sec:hardband}-\ref{sec:nh}), and by performing a fit to the
whole X-ray band with the inclusion of a soft component (Section~\ref{sec:2apec-soft}).

\as1101, also known as S\`{e}rsic~159-03, is located at  R.A.=23h13m58.5s, Dec.= -42d43m39, and the redshift is z=0.058;
for a cosmology of $H_0=73$~km s$^{-1}$ Mpc$^{-1}$, $\Omega_M=0.27$, $\Omega_{\Lambda}=0.73$,
the distance to the cluster is $D_A=220$ Mpc, and 1 arcmin corresponds to 64 kpc.
The paper is structured as follows: in Section~\ref{sec:data} we describe the data reduction methods, in Section~\ref{sec:tprofiles}
we determine the temperature profiles of the clusters from the various observations, 
in Section~\ref{sec:excess} we determine the presence of a soft component above the hot ICM model,
 in Section~\ref{sec:interpretation}
we provide a discussion on the possible origin of the excess emission, and
in Section~\ref{sec:conclusions} we present our conclusions.
Throughout the paper errors are quoted at the 1-$\sigma$, or 68\%  confidence level, unless otherwise
stated.
 
\section{Data and Data Analysis}
In this Section we describe the \chandra, \xmm\ and \rosat\ observations of \as1101, and the method
of analysis of the spatially resolved spectroscopy.
\label{sec:data}
\subsection{\chandra\ observations}
We analyzed \chandra\ observation 11758 of exposure
time 98~ks with the  ACIS-I detector. The data were processed following the method described
in \cite{bonamente2006}, which consists of filtering the observations
for possible periods of flaring background, and applying the latest calibration;
no significant flares were present in this observation.
The reduction was performed in CIAO, using CALDB 4.1.3. The best-fit temperature
of galaxy clusters depends on the effective area and the calibration of ACIS, which has
changed throughout the mission \citep[e.g.,][]{reese2010}, but low-temperature clusters
such as \as1101\ are largely unaffected by the details of the calibration, which impacts
primarily the temperature of clusters at $kT \geq 5$ keV (see e.g. Nevalainen et al., 2010).

The background is measured from blank-sky datasets that are processed following the same
method as the cluster observation \citep{markevitch2003}. 
The blank-sky observations are first rescaled according to the high-energy flux 
of the cluster, to ensure a correct subtraction of the particle background that is dominant
at $E > 9.5$ keV, where the Chandra detectors have no effective area. 
The spectrum of a peripheral region of the observation, beyond 10 arcmin
of the cluster center, is then accumulated; in this region little cluster emission is expected,
and therefore we can check whether there is an enhancement of the soft X-ray background
which could be in principle mistaken for cluster emission. In this observation we measure
no detectable soft X-ray background enhancement, and therefore no additional processing
of the blank-sky background is needed.

\subsection{\xmm\ observations}
We analyzed the pn and MOS data of two \xmm\ observations, 0123900101 and 0147800101.
 In both observations, a thin optical 
blocking filter was used. We processed the raw \xmm\ data with the SASv10.0.0 tools \textit{epchain} and 
\textit{emchain} with the default parameters in order to produce the event files. We used the latest 
calibration information as of September 2010. We also generated the simulated out-of-time event file, 
which we later used to subtract the events from the pn spectra registered during the readout of a CCD. 
We filtered the event files excluding bad pixels and CCD gaps. We further filtered the event files 
including only patterns 0--4 (pn) and 0--12 (MOS). To minimize the contamination of solar particle flares, we
used the $>$10 keV ($>$ 9.5 keV) light curves for the pn (MOS) to accept data only 
from such periods when the count rate is within $\pm$20\% of the quiescent level.
The resulting total exposure time is 85 ks and 222 ks for the pn and combined MOS1+MOS2. 

We used the \textit{evselect-3.60.6} tool to extract spectra, images, and light curves, while excluding the 
regions contaminated by bright point sources. We used the \textit{rmfgen-1.55.2} and \textit{arfgen-1.77.2} tools to 
produce the energy redistribution files and the effective area files. When running the \textit{arfgen} tool, 
we used an extended source configuration and supplied an \xmm\ image of the cluster in detector 
coordinates, binned in 0.5 arcmin pixels, for weighting the response.

We used the blank sky--based estimates for the total sky+particle background spectra from 
Nevalainen et al. (2005).  In order to account for the variability of the instrumental background due 
to cosmic rays, we used the sample of EPIC exposures taken with a \textit{CLOSED} filter (Nevalainen et al. 2005) 
to extract particle background spectra at the same detector regions as used for the cluster data. 
We included this additional component in the fits after adjusting its normalisation so that the 
total background count rate prediction in the 10--14 keV (9.5--12 keV) band for the pn (MOS) matches 
that in the cluster observation.

\subsection{\rosat\ observations}

We analyzed \rosat\ observation 800397 with the Position
Sensitive Proportional Counter (PSPC) detector, for 13~ks of exposure. Data reduction
was performed with XSELECT and the FTOOLS, following the prescriptions
of \cite{snowden1994}, also  described in
\cite{bonamente2002}. In particular we removed periods of high particle
background (master veto rate $\leq$ 170 counts s$^{-1}$) to improve the S/N of the
data.
Given the large field of view ($\sim$ 60 arcmin radius), the PSPC provides a local background for this
observation of \as1101, which is the most accurate for the spectral analysis.
In particular, the local background estimate, obtained simultaneously with the cluster data, enables 
accurate removal of the time-variable Solar wind charge exchange contamination \citep[e.g.,][]{wargelin2004}.

The PSPC detector has a unique feature that makes it especially suitable for the 
detection of soft X-ray radiation from clusters: a large ($\sim$ 200 cm$^2$) and well calibrated 
effective area below the Carbon edge ($\sim$ 0.28 keV). The R2 band, composed of pulse-invariant
channels 20-41 and sometimes referred to as the 1/4~keV band, is sensitive to photons
in the $\sim$~0.2--0.4~keV band, and is the band of choice
to  check for the presence of the soft excess emission in \as1101.
This band is nominally present also in \chandra/ACIS and \xmm/MOS and PN, but
calibrations uncertainties and low effective area prevent their scientific use in these two instruments.

\subsection{Data analysis}
Spectra were accumulated in annuli spaced by 2 arcmin radius, to match the
resolution of the \rosat\ \pspc\ which has an approximate resolution of $\leq$1 arcmin
at all energies. 
We model the spectra using the APEC thermal model \citep{smith2001},
with Wisconsin Galactic absorption cross-sections \citep{morrison1983} and
\cite{anders1989} abundances; in Section~\ref{sec:excess} we present results using
an alternative set of cross-sections \citep{balucinska1992}, 
showing that the choice of cross-sections is not
crucial for the determination of soft X-ray fluxes. 
For \chandra\ and \xmm\ spectra we  let the abundance be a free parameter; for \rosat\
we fix the abundances as explained in Sections~\ref{sec:tprofiles} and \ref{sec:excess}.
For \chandra\ and \xmm\ data we only use the 0.7-7 keV band, and avoid the use of their
lowest-energy channels which are more significantly affected by calibration
uncertainties, and higher background. For \rosat, we use the 0.2-2 keV band (PI channels
20-201), as described in \cite{bonamente2002}.

\section{Spectral fits and 
bandpass dependence of the temperature profile}
\label{sec:tprofiles}
Presence of a soft component in the spectra of galaxy clusters
will result in a bandpass dependence of the best-fit temperature, with
softer bands returning a lower best-fit temperature than high-energy bands 
\citep{cavagnolo2008,lehto2010}. Also, an energy dependent 
calibration problem can produce a similar effect \citep{nevalainen2010}.

\begin{figure}
\centering
\includegraphics[width=1.7in,angle=-90]{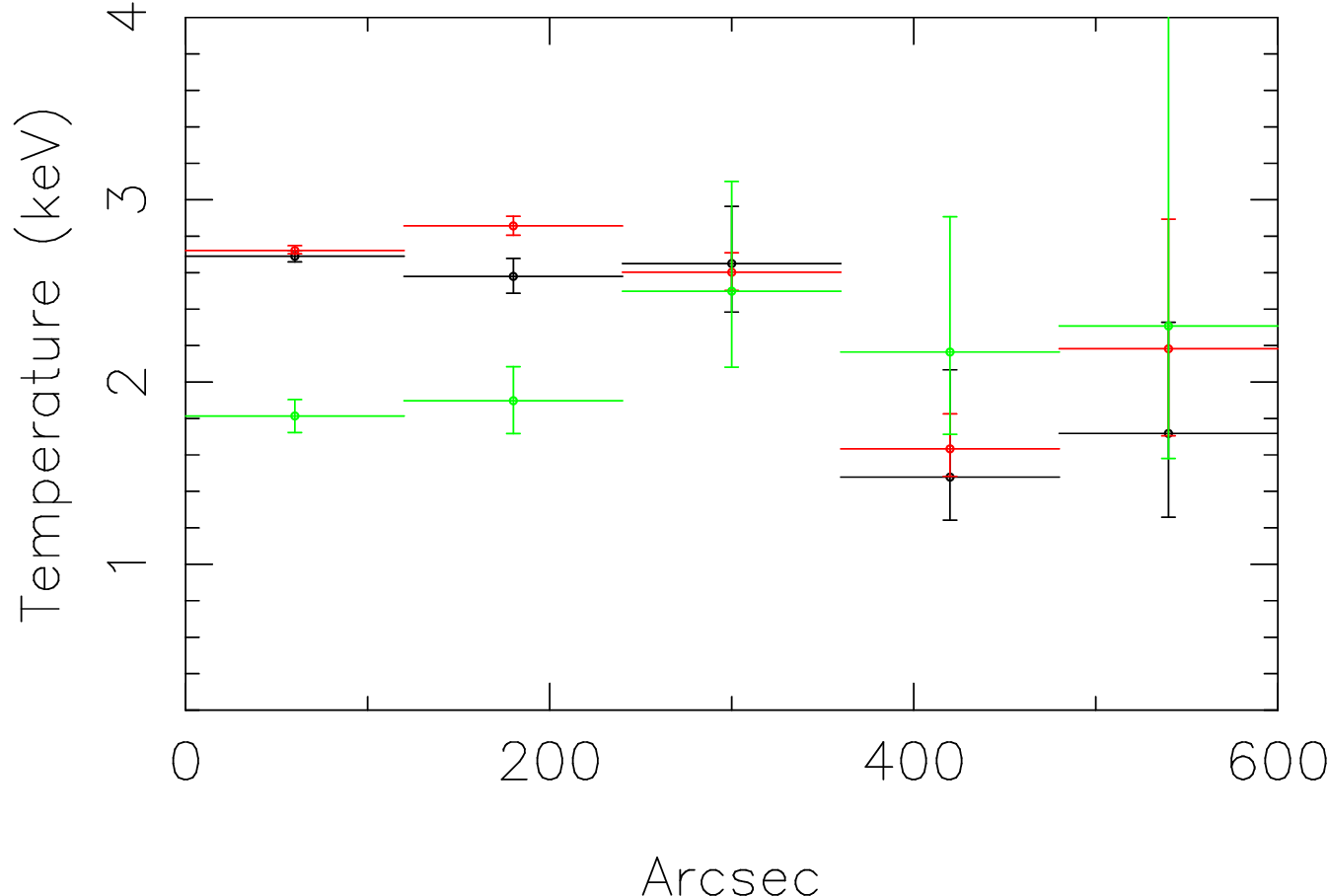}
\includegraphics[width=1.7in,angle=-90]{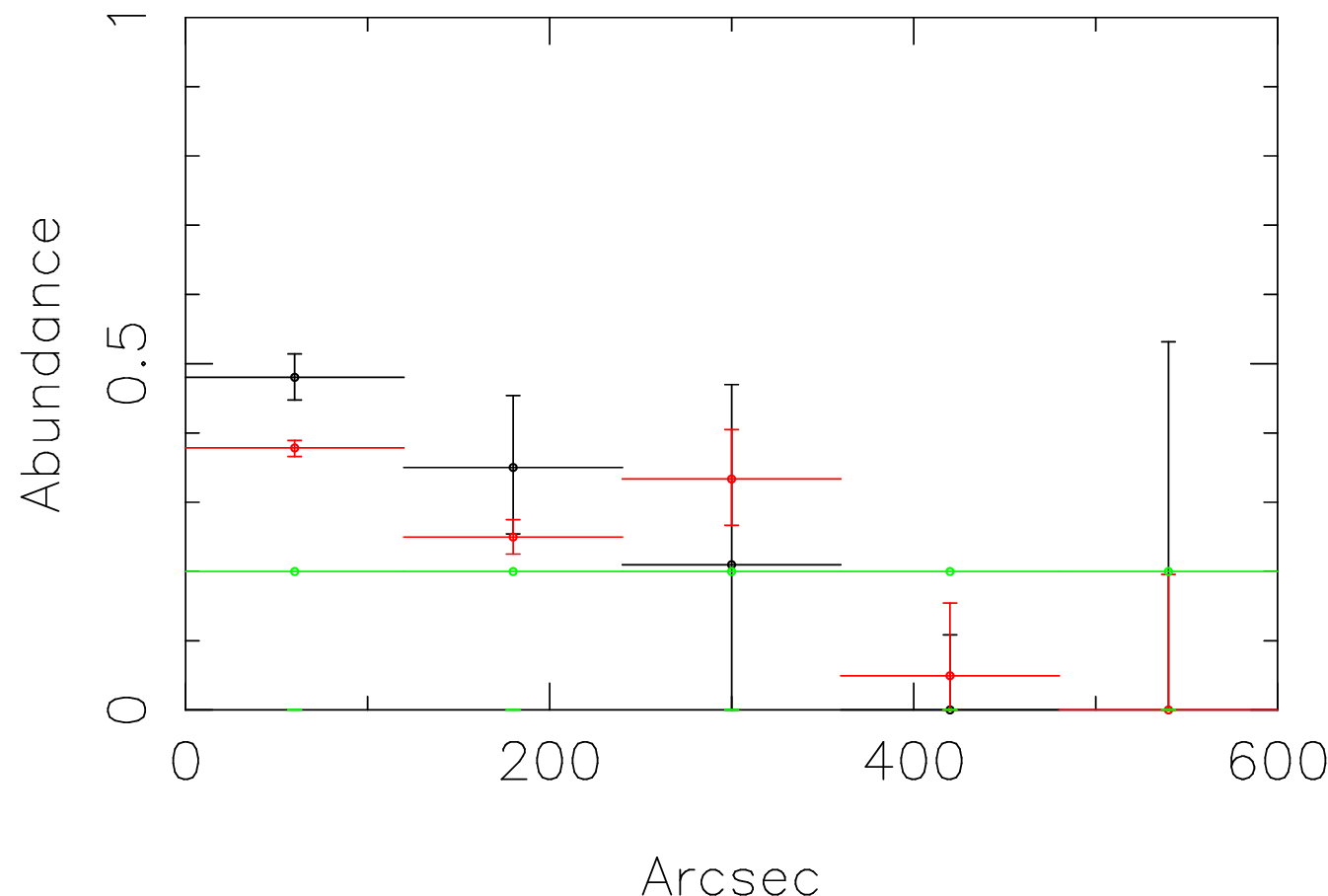}
\caption{Temperature and abundance profiles in hard X-ray band (2-7 keV).
Black: \chandra; Red: \xmm. Overplotted are also the whole band  (0.2-2 keV) 
results for \rosat\ (green).
}
\label{fig:t-hard-chandra-xmm-rosat}
\end{figure}

To address the calibration of \chandra\ and \xmm, we first fitted the \chandra\  and \xmm\ spectra
in the wide band (0.7-7.0 keV)  and hard band (2.0-7.0 keV), using a single-temperature APEC model. 
We found that in the hard band the temperatures are consistent between the two instruments,
as shown in Tables~\ref{tab:tprofiles-1}
and \ref{tab:ratios}, and Figure~\ref{fig:t-hard-chandra-xmm-rosat}
(the largest difference is at less than the 3-$\sigma$ level in one bin, for a ratio between \chandra\ and 
\xmm\ temperatures of 0.903$\pm$0.037).
Agreement of hard-band temperatures was also
 found for a sample of clusters observed with  \chandra\ and \xmm\ 
by \cite{nevalainen2010}, where
it was shown that the bremsstrahlung temperatures agree with the 
Fe XXV/XXVI line ratio temperatures. It is therefore
likely that the hard band calibration of \xmm\ and \chandra\ 
does not have significant energy-dependent problems.

\begin{figure}
\centering
\includegraphics[width=1.7in,angle=-90]{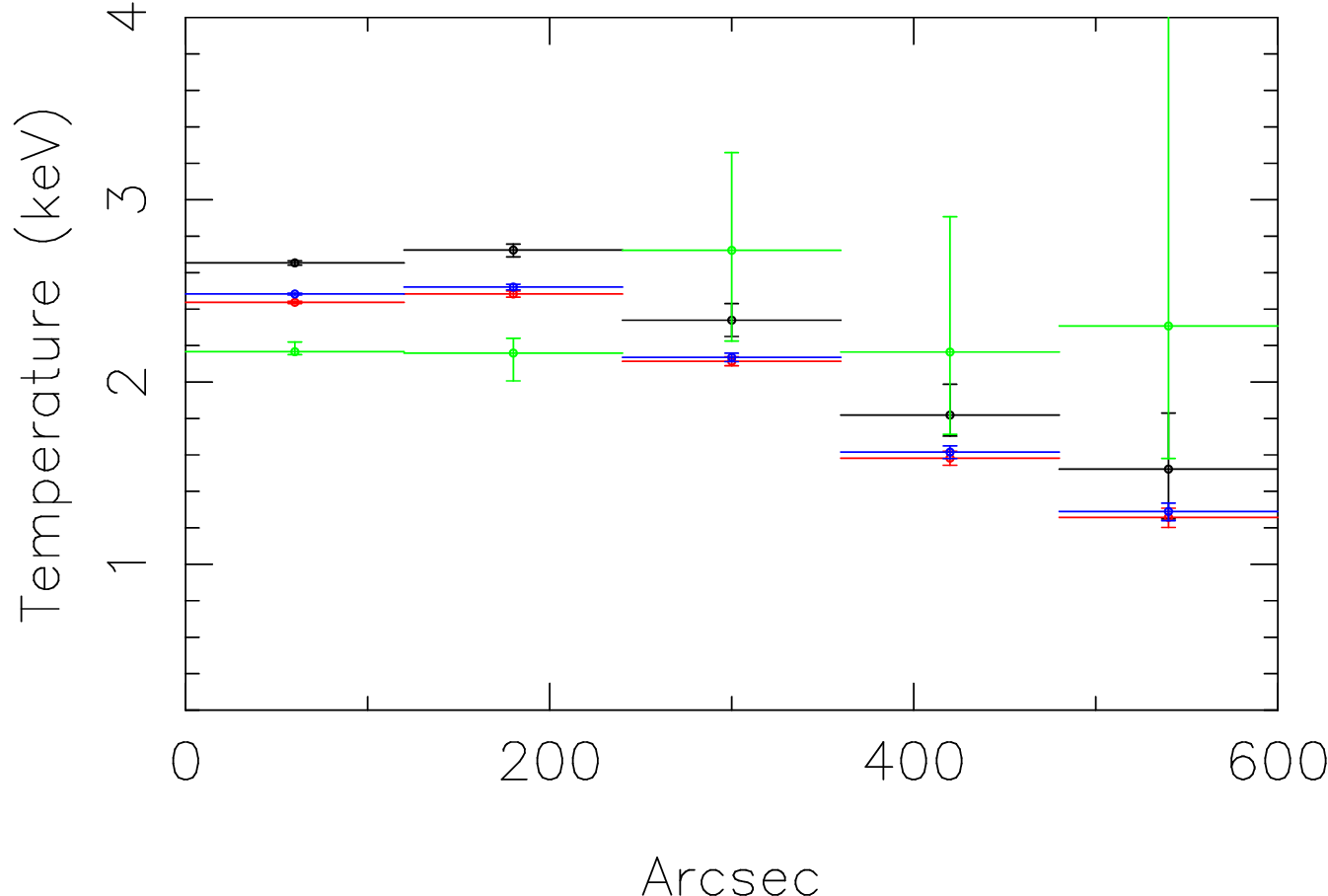}
\includegraphics[width=1.7in,angle=-90]{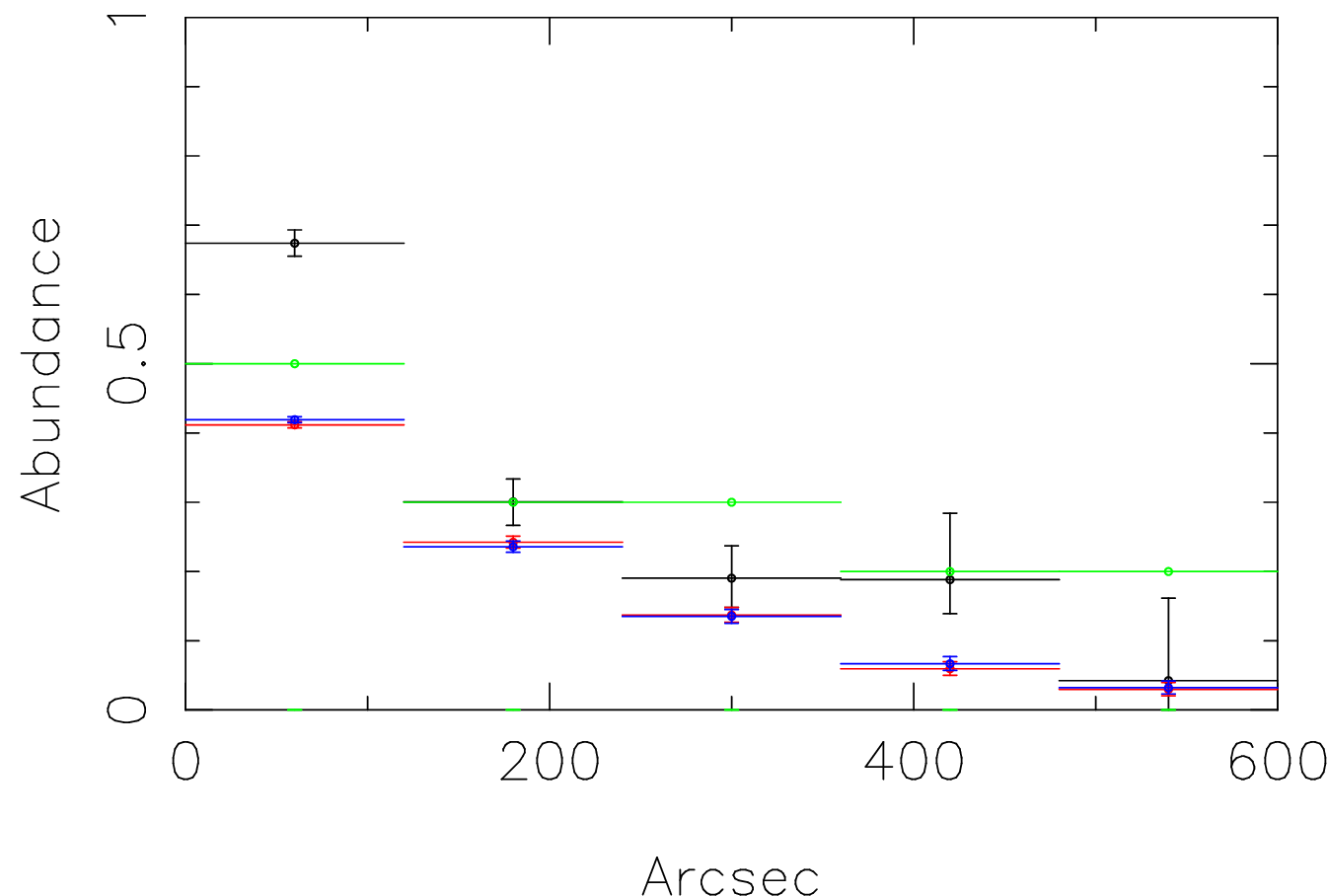}
\caption{Temperature and abundance profiles in whole X-ray band.
Black: \chandra\ (0.7-7 keV); Red: \xmm\ (0.7-7 keV); Green: ROSAT (0.2-2 keV);
blue: all instruments combined.
}
\label{fig:t-whole}
\end{figure}

In the wide band, however, the temperatures obtained using 
\xmm\ are systematically and significantly 
lower than those  obtained with \chandra\ (see Figure~\ref{fig:t-whole} and 
Tables~\ref{tab:tprofiles-1} and \ref{tab:ratios}).
Again, this is consistent with the analysis of the \xmm\ - \chandra\ sample  \citep{nevalainen2010} and
is likely the result of remaining calibration uncertainties in 
the energy band $\leq$ 2~keV for any of the detectors.

\begin{figure}
\centering
\includegraphics[width=1.7in,angle=-90]{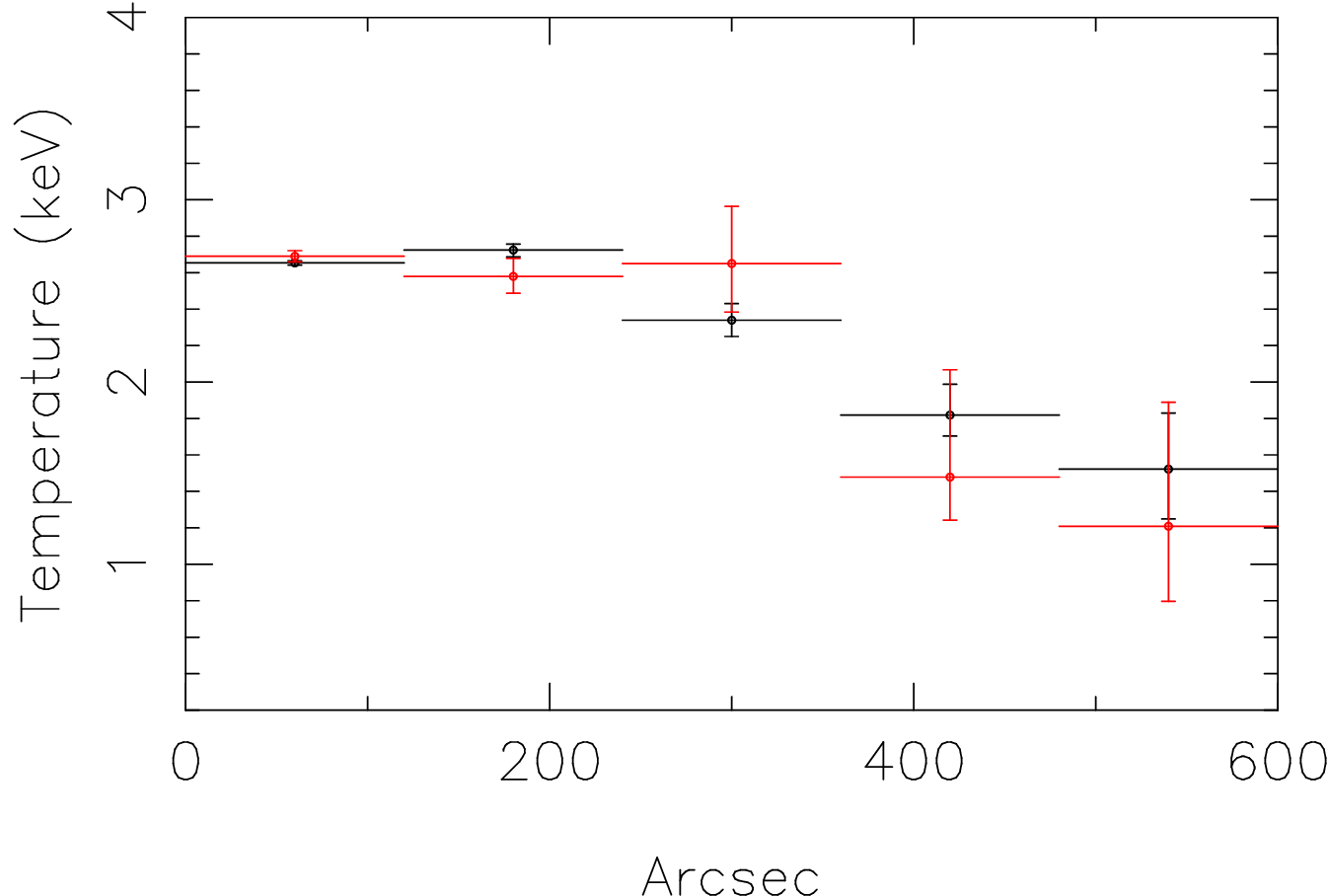}
\includegraphics[width=1.7in,angle=-90]{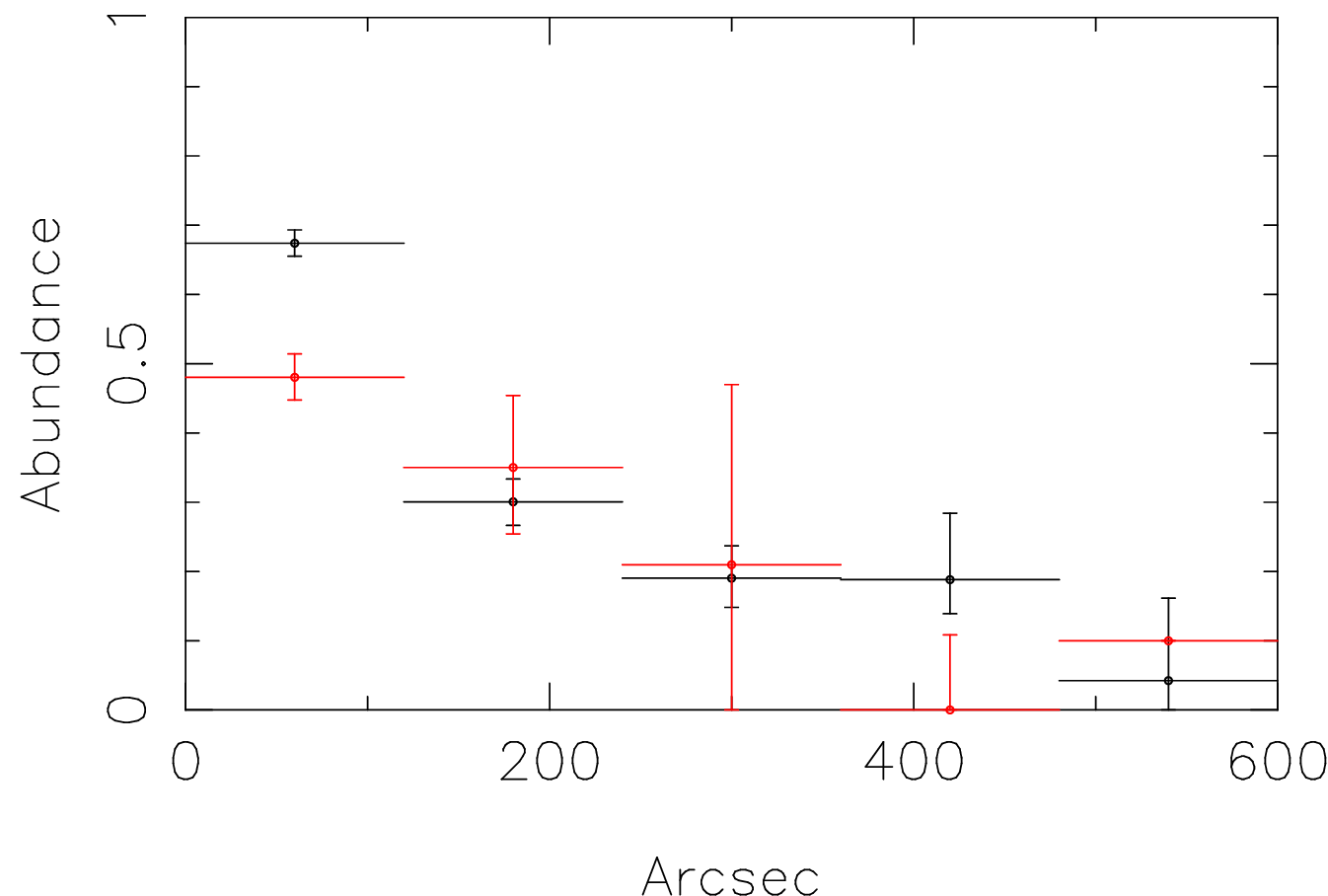}
\caption{\chandra\ temperature and abundance profiles in whole X-ray band (black), compared with profiles in
hard band (red, 2-7 keV).}
\label{t-chandra}
\end{figure}

\begin{figure}
\centering
\includegraphics[width=1.7in,angle=-90]{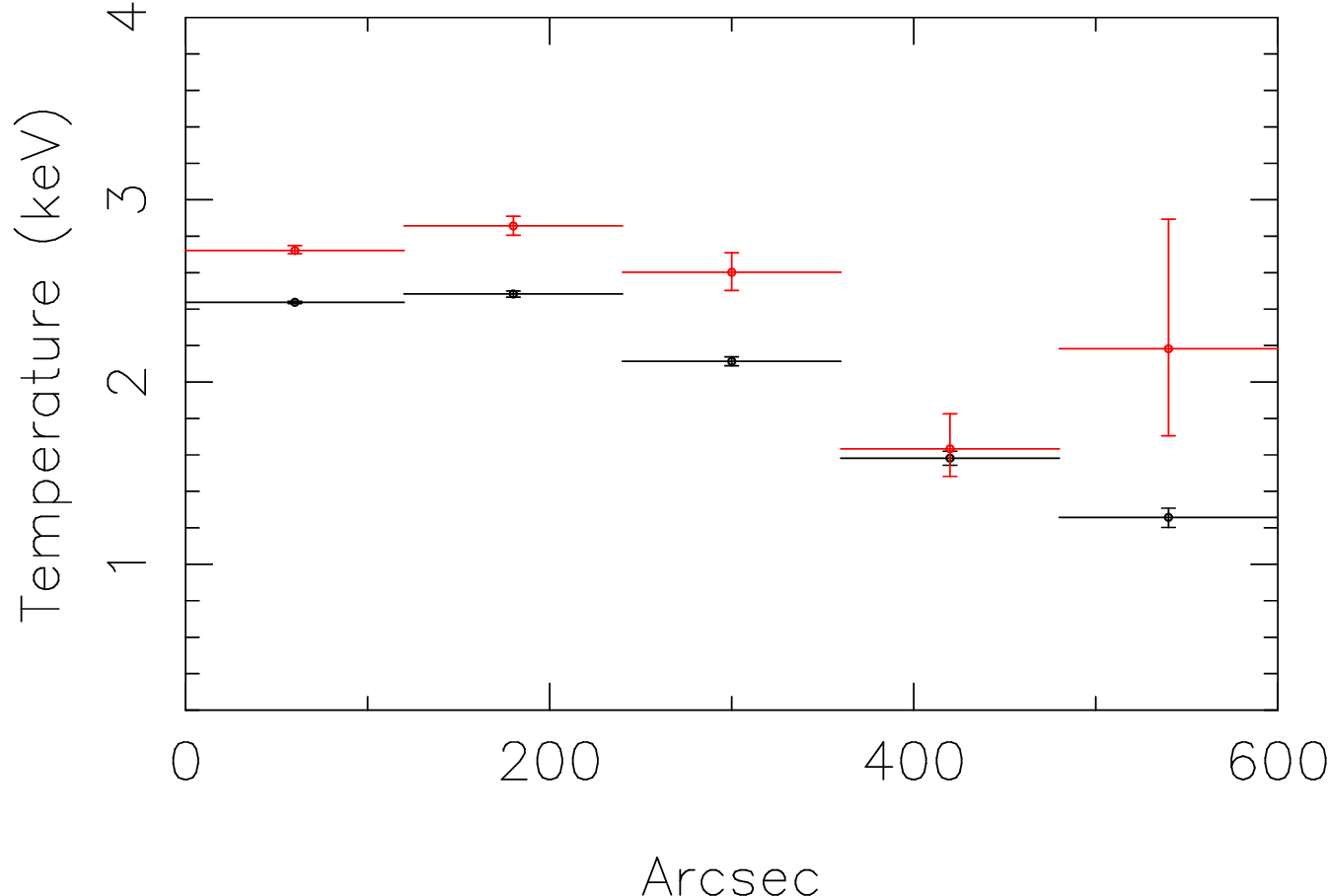}
\includegraphics[width=1.7in,angle=-90]{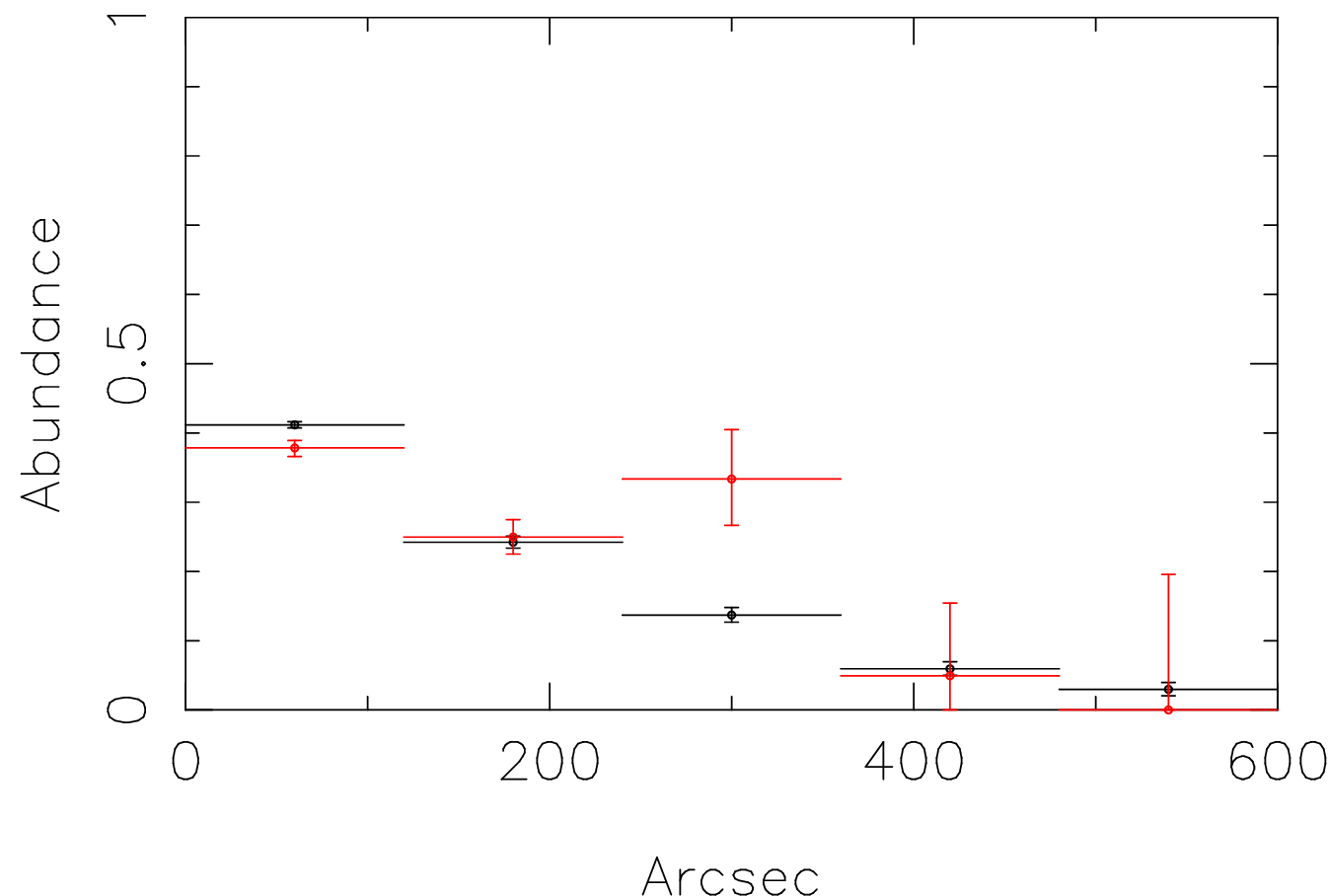}
\caption{\xmm\ temperature and abundance profiles in whole X-ray band (black), compared with profiles in
hard band (red, 2-7 keV).}
\label{t-xmm}
\end{figure}

Thus the comparison of the hard and wide band temperatures yields
 different results for \chandra\  and \xmm:
the best-fit \chandra\ temperatures in the whole X-ray band
and in the hard band are statistically consistent at all radii, indicating that there is no discernible
soft X-ray excess component at these energies (see Fig.~\ref{t-chandra}).
The \xmm\ results differ from the \chandra\ results in that the
wide-band fit detects a significantly lower temperature than the hard band (Figure~\ref{t-xmm}), 
especially in the regions of highest S/N, consistent with
the presence of soft excess in this cluster, as reported by \cite{bonamente2005}. 
Thus, with the current calibration we cannot establish whether the 0.7-2 keV band has evidence 
for soft excess emission in both \chandra\ and \xmm.


\rosat\ wide band (0.2-2 keV) temperatures are significantly  lower than the 
hard band temperatures of \chandra\ and \xmm\ in the inner two regions, where
the signal has the highest S/N.
(see Figure~\ref{fig:t-hard-chandra-xmm-rosat}  and Tables \ref{tab:tprofiles-1}
and \ref{tab:tprofiles-2}).
Assuming that the \rosat\ calibration is correct, with effective
area calibration errors at less than 5\% \citep[e.g.][]{snowden1994,beuermann2008},
this difference can be explained by the existence of an additional soft component 
emitting primarily in the \rosat\ band, indicating that the soft excess is dominant at energies
below 1 keV.
The best-fit \rosat\ wide band temperatures are consistent with those from
\chandra\ and \xmm\ in the other regions where the S/N is lower.



We also study the dependence of the \rosat\ temperature profiles on the assumed abundance profile, given that
we cannot measure it directly from the \rosat\ data themselves. 
We find that the two abundance profiles
used (a decreasing one and a constant one), have negligible effect on the best-fit temperatures,
especially given the statistical quality of the data (see Table~\ref{tab:tprofiles-2}).


%
%

\begin{table}
\footnotesize
\centering
\begin{tabular}{lccc}
\hline
Annulus  & \multicolumn{3}{c}{} \\
\hline
 & \multicolumn{3}{c}{\chandra\ 0.7-7 keV fit} \\
(arcmin)  & $kT$ (keV) & $A$ & $\chi^2$ (d.o.f) \\
\hline
0-2 & 2.65 $\pm^{0.01}_{0.01}$ & 0.67 $\pm^{0.02}_{0.02}$ &  647.12 (413)\\
2-4 & 2.72 $\pm^{0.03}_{0.04}$ & 0.30 $\pm^{0.03}_{0.03}$ &  445.35 (422)\\
4-6 & 2.34 $\pm^{0.09}_{0.09}$ & 0.19 $\pm^{0.05}_{0.04}$ &  375.57 (429)\\
6-8 & 1.82 $\pm^{0.17}_{0.12}$ & 0.19 $\pm^{0.10}_{0.05}$ &  475.17 (429)\\
8-10&1.52 $\pm^{0.31}_{0.27}$ & 0.04 $\pm^{0.12}_{0.04}$ &  453.47 (420)\\
\hline
 & \multicolumn{3}{c}{\chandra\ 2-7 keV fit}\\
(arcmin)  & $kT$ (keV) & $A$ & $\chi^2$ (d.o.f) \\
\hline
0-2 & 2.69 $\pm^{0.03}_{0.03}$ & 0.48 $\pm^{0.03}_{0.03}$ &  334.16 (324)\\
2-4 & 2.58 $\pm^{0.10}_{0.09}$ & 0.35 $\pm^{0.10}_{0.10}$ &  332.08 (333)\\
4-6 & 2.65 $\pm^{0.31}_{0.27}$ & 0.21 $\pm^{0.26}_{0.21}$ &  305.73 (340)\\
6-8 & 1.48 $\pm^{0.59}_{0.24}$ & 0.00 $\pm^{0.11}_{0.00}$ &  381.04 (340)\\
8-10& 1.21 $\pm^{0.68}_{0.41}$ & 0.10 $\pm^{0.00}_{0.00}$ &  338.31 (331)\\
\hline
 & \multicolumn{3}{c}{\xmm\ 0.7-7 keV fit}\\
(arcmin)  & $kT$ (keV) & $A$ & $\chi^2$ (d.o.f) \\
\hline
0-2 & 2.44 $\pm^{0.01}_{0.01}$ & 0.41 $\pm^{0.00}_{0.00}$ &  2230.33 (1765)\\
2-4 & 2.48 $\pm^{0.02}_{0.02}$ & 0.24 $\pm^{0.01}_{0.01}$ &  974.86 (1043)\\
4-6 & 2.11 $\pm^{0.02}_{0.02}$ & 0.14 $\pm^{0.01}_{0.01}$ &  690.69 (723)\\
6-8 & 1.58 $\pm^{0.04}_{0.04}$ & 0.06 $\pm^{0.01}_{0.01}$ &  554.84 (571)\\
8-10&1.26 $\pm^{0.05}_{0.06}$ & 0.03 $\pm^{0.01}_{0.01}$ &  521.85 (574)\\
\hline
 & \multicolumn{3}{c}{\xmm\ 2-7 keV fit}\\
(arcmin)  & $kT$ (keV) & $A$ & $\chi^2$ (d.o.f) \\
\hline
0-2 & 2.72 $\pm^{0.03}_{0.02}$ & 0.38 $\pm^{0.01}_{0.01}$ &  918.25 (1063)\\
2-4 & 2.86 $\pm^{0.05}_{0.05}$ & 0.25 $\pm^{0.03}_{0.02}$ &  293.64 (429)\\
4-6 & 2.60 $\pm^{0.11}_{0.10}$ & 0.33 $\pm^{0.07}_{0.07}$ &  213.96 (276)\\
6-8 & 1.63 $\pm^{0.19}_{0.15}$ & 0.05 $\pm^{0.11}_{0.05}$ &  255.13 (244)\\
8-10& 2.18 $\pm^{0.71}_{0.48}$ & 0.00 $\pm^{0.20}_{0.00}$ &  203.04 (281)\\
\hline
\end{tabular}
\caption{Best-fit temperature profiles from \chandra\ and \xmm. \label{tab:tprofiles-1}}
\end{table}

\begin{table*}
\footnotesize
\centering
\begin{tabular}{lcccc}
\hline
Annulus  &  \chandra & \xmm  & \chandra/\xmm& \chandra/\xmm \\
(arcmin) &  2-7/0.7-7 keV &  2-7/0.7-7 keV & 0.7-7 keV & 2-7 keV\\
\hline
0-2 & 1.014 $\pm$ 0.013 & 1.117 $\pm$0.009& 1.089 $\pm$0.005 & 0.988 $\pm$0.014 \\
2-4 & 0.947 $\pm$ 0.037 & 1.151 $\pm$0.022& 1.097 $\pm$0.016 & 0.903 $\pm$0.037 \\
4-6 & 1.133 $\pm$0.131  & 1.231 $\pm$0.051& 1.107 $\pm$0.045 & 1.019 $\pm$0.119 \\
6-8 & 0.813 $\pm$0.236  & 1.032 $\pm$0.111& 1.149 $\pm$0.094 & 0.905 $\pm$0.270 \\
8-10& 1.128 $\pm$0.412  & 1.735 $\pm$0.478& 1.210 $\pm$0.236 & 0.787 $\pm$0.326  \\
\hline
\end{tabular}
\caption{Ratio of temperatures from \chandra\ and \xmm\ data. \label{tab:ratios}}
\end{table*}

\begin{table}
\footnotesize
\centering
\begin{tabular}{lccc}
\hline
Annulus  & \multicolumn{3}{c}{} \\
\hline
 & \multicolumn{3}{c}{\rosat\ 0.2-2 keV fits}\\
(arcmin)  & $kT$ (keV) & $A$ & $\chi^2$ (d.o.f) \\
\hline
0-2 & 1.81 $\pm^{0.09}_{0.09}$ & 0.20  &  197.07 (161)\\
2-4 & 1.90 $\pm^{0.19}_{0.18}$ & 0.20  &  115.16 (104)\\
4-6 & 2.50 $\pm^{0.60}_{0.42}$ & 0.20  &  69.01 (60)\\
6-8 & 2.16 $\pm^{0.74}_{0.45}$ & 0.20  &  52.58 (47)\\
8-10& 2.31 $\pm^{7.93}_{0.73}$ & 0.20  &  59.74 (45)\\
\hline
0-2 & 2.17 $\pm^{0.05}_{0.02}$ & 0.50  &  250.97 (161)\\
2-4 & 2.16 $\pm^{0.08}_{0.15}$ & 0.30  &  123.93 (104)\\
4-6 & 2.72 $\pm^{0.54}_{0.50}$ & 0.30  &  70.92 (60)\\
6-8 & 2.16 $\pm^{0.74}_{0.45}$ & 0.20  &  52.58 (47)\\
8-10& 2.31 $\pm^{7.93}_{0.73}$ & 0.20  &  59.74 (45)\\
\hline
\end{tabular}
\caption{Best-fit temperature profiles from \rosat.\label{tab:tprofiles-2}}
\end{table}

\section{Measurement of soft excess fluxes}
\label{sec:excess}
Since we are unable to determine conclusively 
whether there is excess emission in the \chandra\ or \xmm\ 0.7-2 keV band
of these \as1101\ observations,
and given the evidence for an
  additional emission component in the lowest \rosat\ channels provided in
  Section~\ref{sec:tprofiles}, we proceed to the analysis of the soft excess in the 1/4
  keV ROSAT band (R2, 0.2-0.4 keV).
In Section~\ref{sec:hardband} we calculate soft excess fluxes using an extrapolation 
from the hard band model, following a procedure we used in earlier papers
\citep[e.g.][]{bonamente2005,nevalainen2003,bonamente2002}. We then calculate
the soft excess from fits that include the entire 0.7-7 keV \chandra/\xmm\ 
bandpasses (Section~\ref{sec:wholeband}),
and confirm the presence of soft excess emission. The physical characteristics of the
soft emitter are determined in Sections~\ref{sec:2apec-soft} and interpreted
in Section~\ref{sec:interpretation}, 
by modelling the whole-band  \chandra, \xmm\ and \rosat\ spectra  
with two components that simultaneously 
describe both the hot ICM emission and the soft excess emitter.

\subsection{Soft X-ray fluxes from hard-band model}
\label{sec:hardband}
\begin{figure}
\centering
\includegraphics[width=1.7in,angle=-90]{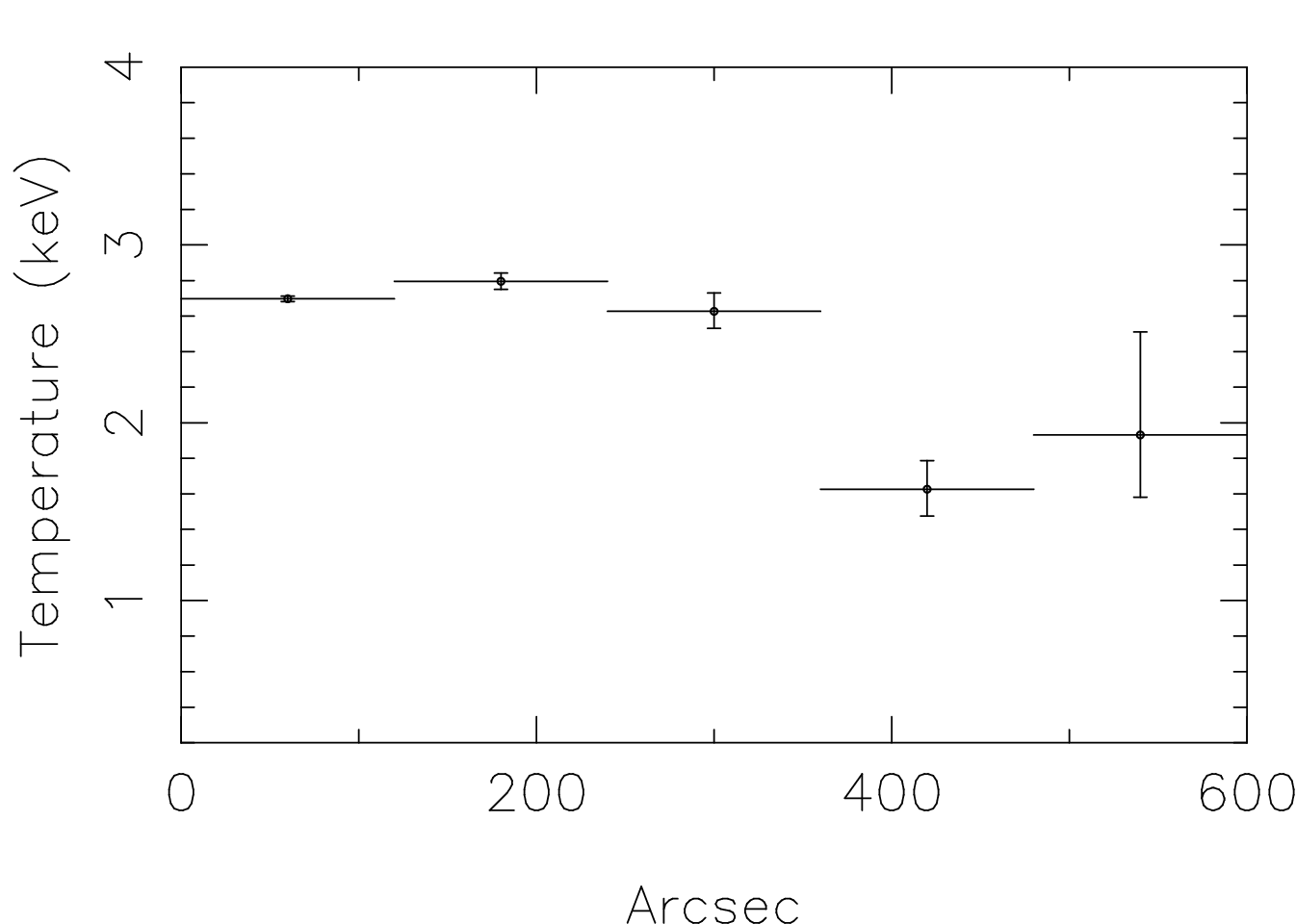}
\includegraphics[width=1.7in,angle=-90]{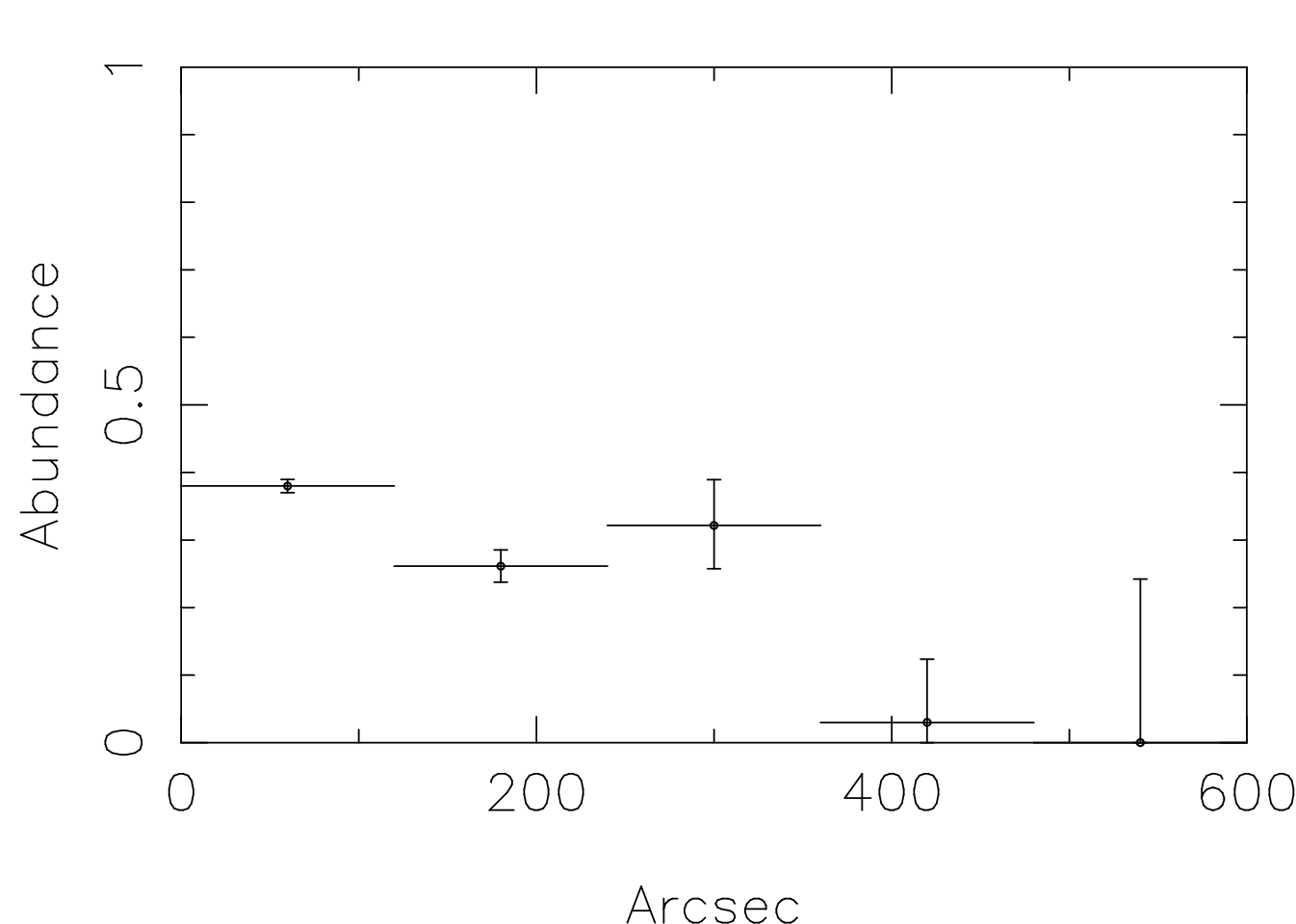}
\caption{Temperature and abundance profile in high-energy band from 
composite \chandra, \xmm\ and \rosat\ high energy-data, using a single temperature model
for the hot gas.
}
\label{fig:t-hard}
\end{figure}

\begin{figure}
\centering
\includegraphics[width=2in,angle=-90]{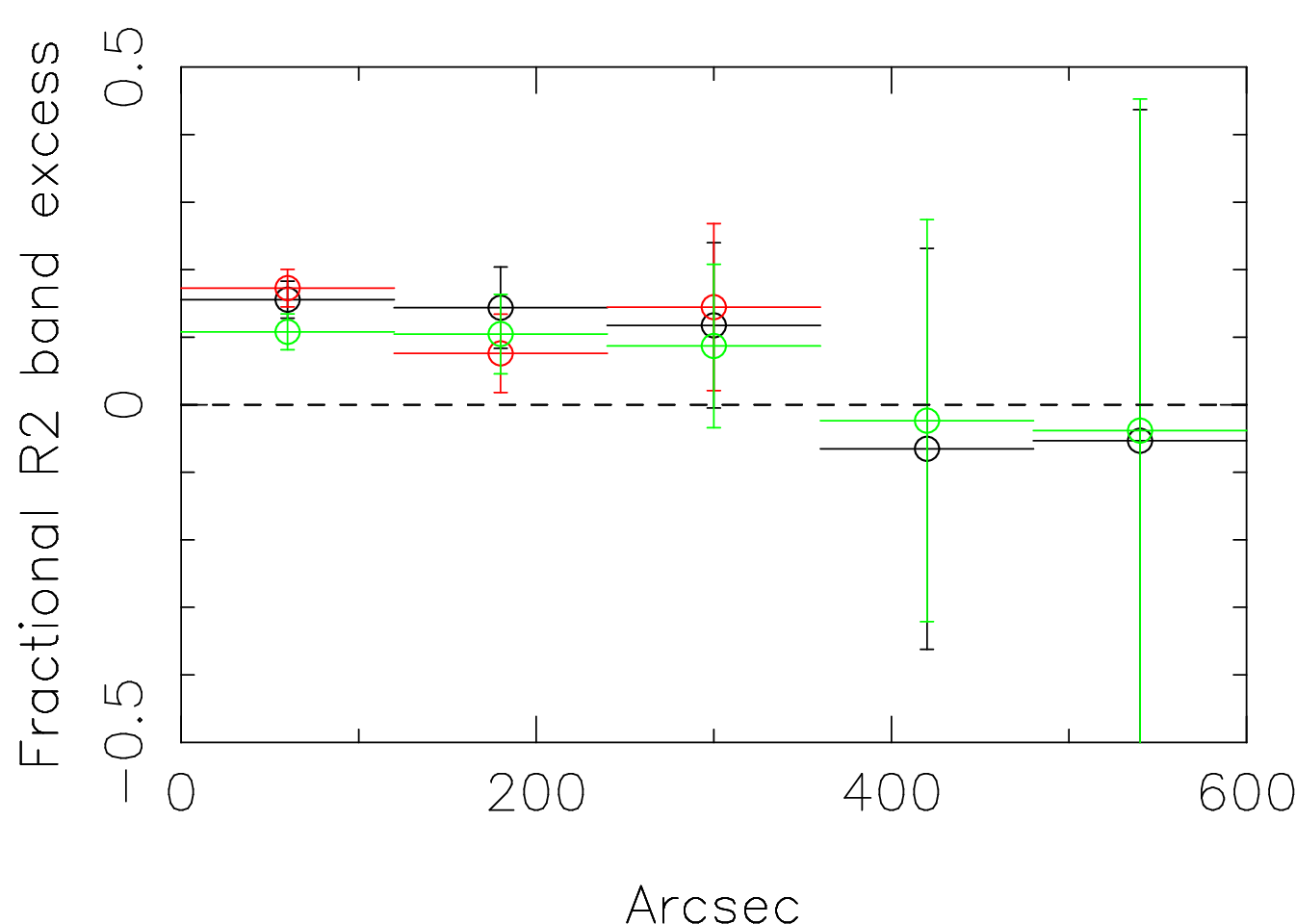}
\includegraphics[width=2in,angle=-90]{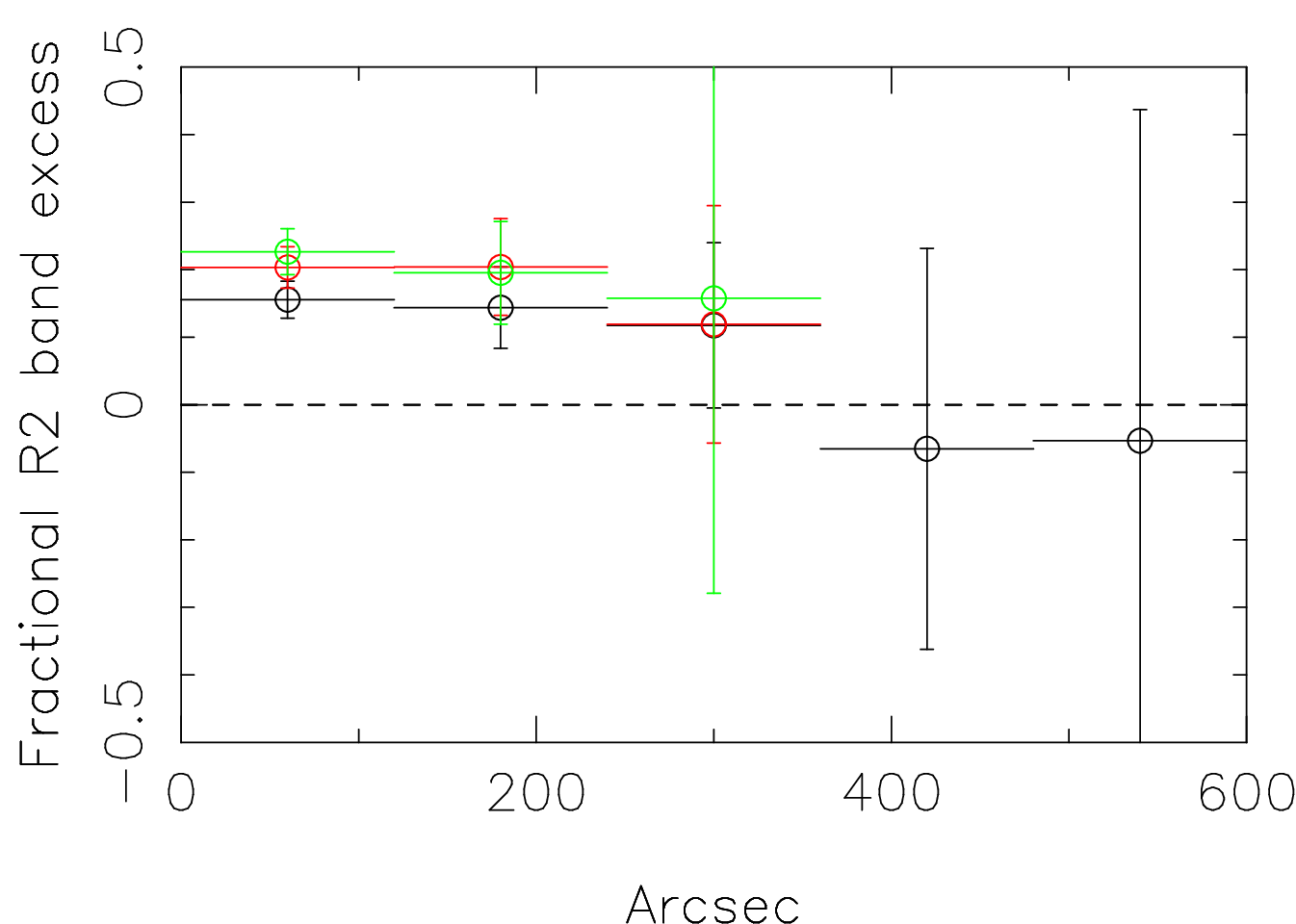}
\caption{
(Left) R2 band soft residual profile from  high-energy 
composite \chandra, \xmm\ and \rosat\ fits. 
Black: fit is to a single APEC model
with  \cite{morrison1983} cross-sections; Red: fit is to a two-APEC model
with \cite{morrison1983} cross-sections; Green: fit is to a single APEC model
with  \cite{balucinska1992}  cross-sections.
(Right) R2 band soft residual profile from whole-band 
fits to \chandra\ and \rosat\ data (green) and \xmm\ and \rosat\ data (red); 
for comparison, in black
are the residuals from the fit is to a single APEC model, as in left panel.}
\label{fig:soft-excess}
\end{figure}

Given the evidence presented in Section~\ref{sec:tprofiles} that the hard band 
calibration of \chandra\ and \xmm\ is accurate,
 we proceed with
obtaining a joint hard band temperature profile. 
This is accomplished by fitting simultaneously the 2-7~keV band data from 
\chandra\ and \xmm\, with the addition of  the 1-2~keV band data from \rosat.
We find that the normalization constant of the \chandra, \xmm\ and \rosat\
best-fit models, after taking into account exclusion regions, are consistent within 10\% for all instruments.
This result is consistent with the current estimate of the \xmm/\chandra\ flux
cross-calibration uncertainty of
$\sim$10\% \citep{nevalainen2010}.
Results are presented in Figure~\ref{fig:t-hard} and  Table~\ref{tab:data},
where we also report the percentage of total measured 
R2 band fluxes  above the background, as measured from the same \rosat\ observations.
%
%
We extrapolate this temperature profile to the \rosat\ R2 band, and determine the
fractional soft excess flux as
\begin{equation*}
\eta = \frac{\text{flux}(R2)-\text{model}(R2)}{\text{model}(R2)}
\end{equation*}
as defined in \cite{bonamente2002}.
Uncertainties in the hard-band models are included in the calculation of the
soft excess fluxes.
The radial profile of the fractional R2 band emission indicates a low level ($\sim$10--20\%)
of excess emission at radii $\leq$ 6 arcmin, and no soft residuals beyond that radius
(Figure~\ref{fig:soft-excess}).
The fact that the soft emission at large radii is consistent with the hot gas prediction
is also an indication that the Galactic soft X-ray background was properly subtracted.

The left-hand panel of Figure~\ref{fig:spectra} shows the \chandra, \xmm\ and \rosat\ 
spectra in the 0-2 and  2-4 arcmin regions,
with the best-fit model obtained by a fit to the hard band of the three instruments.
The spectra illustrate three points: at $E>2$~keV for \chandra\ and \xmm, and $E>1$ keV for \rosat,
the three instruments are consistent; at $E=0.7-2$~keV, \chandra\ has no evidence for an excess
emission while both \xmm\ instruments do; and,  in the softest  band (0.2-0.4 keV), \rosat\ appears to
detect an excess of emission. Similar conclusions apply to the 4-6 arcmin region.

To test for the impact of cooler gas in the core, or the effect of projection of cooler gas from
outer regions, we also use a two-temperature APEC model (see Table~\ref{tab:data-2apec}) in  
which the second thermal model is constrained to have the same abundance 
and half the temperature of the first model, as done also by \cite{kaastra2003}.
In fact, we know from X-ray spectroscopy that gas in the core is not found to cool
much below half the peak temperature \citep{tamura2001,peterson2003}.
The exercise shows that the soft residuals are still present when using this multi-phase model
(see Fig.~\ref{fig:soft-excess}, red curve in left panel).
Therefore we conclude that cooler gas with temperatures within one half that of the virial
gas cannot explain the soft X-ray radiation in the 1/4 keV band.

We also test for the effect of the cross-section of Galactic absorbers in the derivation of the
soft excess fluxes. Throughout the paper, we used the cross-sections of \cite{morrison1983}
with the abundances of \cite{anders1989}.
Of particular relevance for the study of the soft excess is the cross-section of helium,
which has been revised several times, for example by \cite{yan1998}.
Recently the \cite{balucinska1992} cross-sections
were also revised,
and we test the presence of the soft excess by use of these cross-sections, in conjunction with
the \cite{grevesse1998} abundances.
We fit the joint X-ray spectra to a single-temperature thermal model, and 
the results presented in Table~\ref{tab:data-phabs} show that the \cite{morrison1983} and
 the  \cite{balucinska1992} models
differ by less than 5\% in the 1/4 keV band, and the excess is  detected with high significance in both cases
(see Fig.~\ref{fig:soft-excess}, green curve in left panel). 
We therefore conclude that uncertainties in the cross-section of absorbing gas is not a significant factor
in the determination of soft X-ray fluxes.

\subsection{Soft X-ray fluxes from whole-band  fits}
\label{sec:wholeband}
We also determine the soft excess flux in the R2 band when the hot ICM is determined from 
a whole-band fit. Given that the \chandra\ and \xmm\ instrument have systematic
differences in the 0.7-2 keV fluxes, we perform two separate fits:
one to the 0.7-7~keV \chandra\ plus 0.5-2~keV \rosat\ data, and one to the
0.7-7~keV \xmm\ plus 0.5-2~keV \rosat\ data.
Results are presented in Table~\ref{tab:data-chandra-xmm-rosat-whole} and in Figure~\ref{fig:soft-excess} (right panel), 
and are qualitatively similar to those obtained from the hard-band model:
soft excess emission in the R2 band is detected for the two inner annuli (0-4 arcmin)
with a confidence $\geq 2 \sigma$. 
These results provide further evidence for the presence
of soft excess emission in the R2 band above the contribution from the hot ICM, regardless
of calibration uncertainties in the 0.7-2 keV band between \chandra\ and \xmm.

\subsection{Free-$N_H$ fits to the whole-band \chandra, \xmm\ and \pspc\ data}
\label{sec:nh}
Another possibility for the appearance of soft residual fluxes in the R2 band is the
use of an incorrect Galactic column density. We therefore perform a fit to the
combined \chandra, \xmm\ and \rosat\ spectra in the entire spectral range, including
the R2 band for \rosat, and show the results of our fit in Table~\ref{tab:data-2apec-freeNH}.
These fits result in no soft excess fluxes for all regions in the R2 band, and indicate
that the Galactic $N_H$ required to explain the \rosat\ soft fluxes must be in the range
$N_H=7-9 \times 10^{20}$ cm$^{-2}$, or approximately 30\% lower than the value
of 1.15 $ \times 10^{20}$ cm$^{-2}$ measured by the \cite{kalberla2005} survey.
Based on \xmm\ and \suzaku\ observation, \cite{werner2007} finds that the best-fit
$N_H$ towards the central region of the cluster is consistent with the
LAB value, i.e., there is no need for soft excess component there, and that
the excess emission remains present at large radii when using the LAB $N_H$ value
\citep{durret2008}. The difference between these results and those of \cite{werner2007}  for the innermost
 region may be explained
with the use of the \rosat\ data covering the 1/4~keV band.

\subsection{Fit to  whole-band \chandra, \xmm\ and \pspc\ data with an additional
soft model}
\label{sec:2apec-soft}
\begin{figure*}
\centering
\includegraphics[width=2.2in,angle=-90]{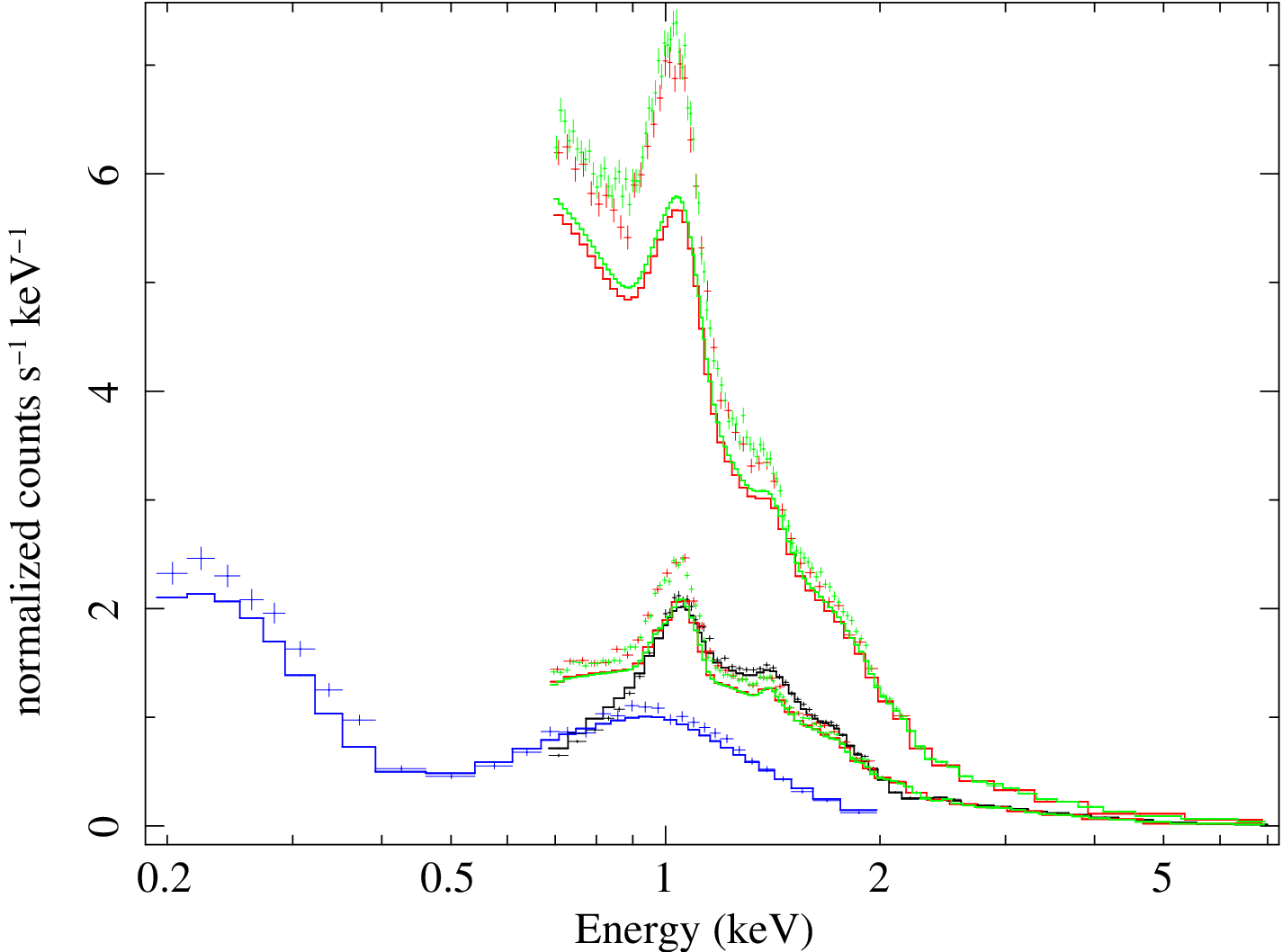}
\includegraphics[width=2.2in,angle=-90]{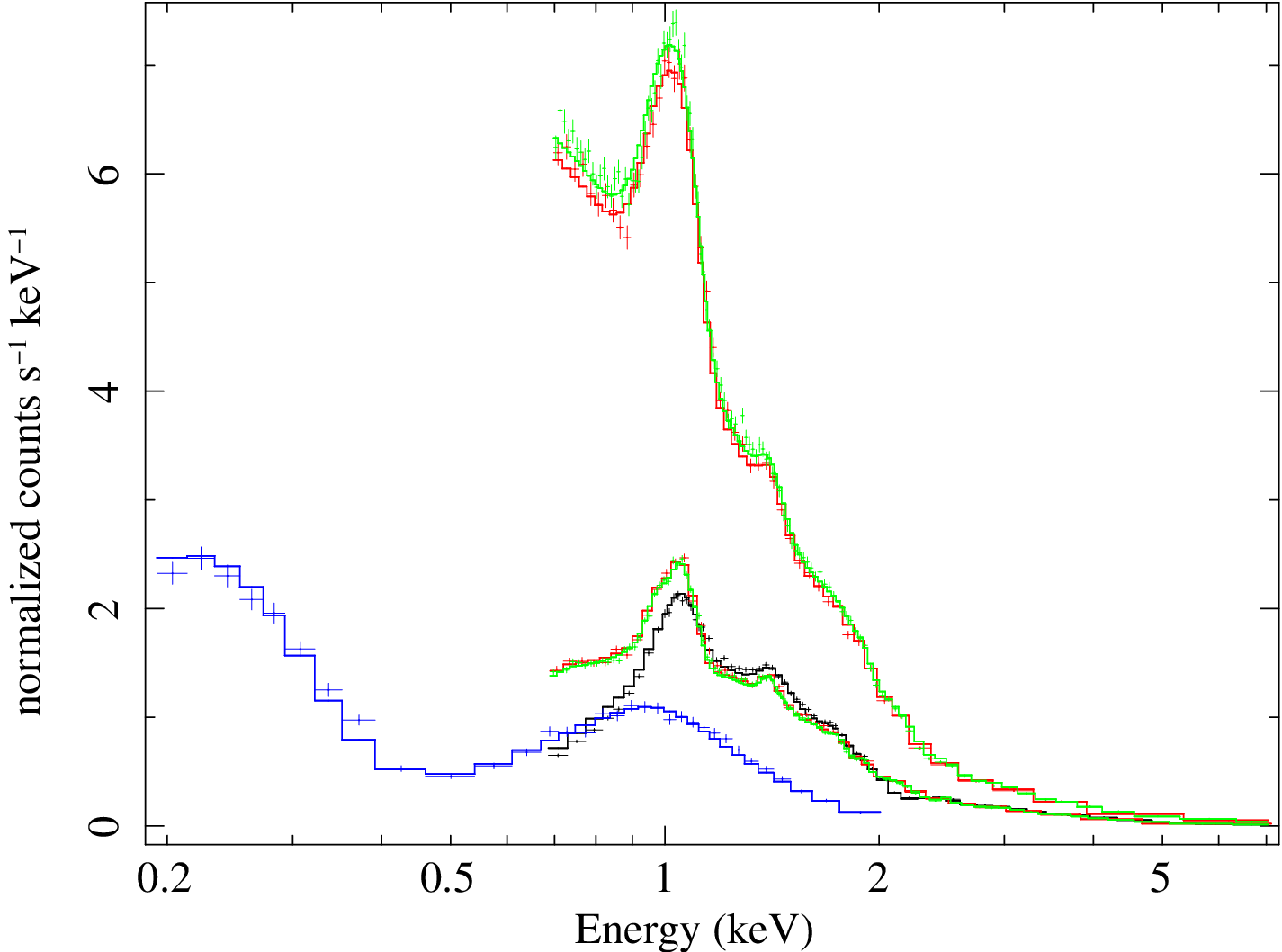}

\includegraphics[width=2.2in,angle=-90]{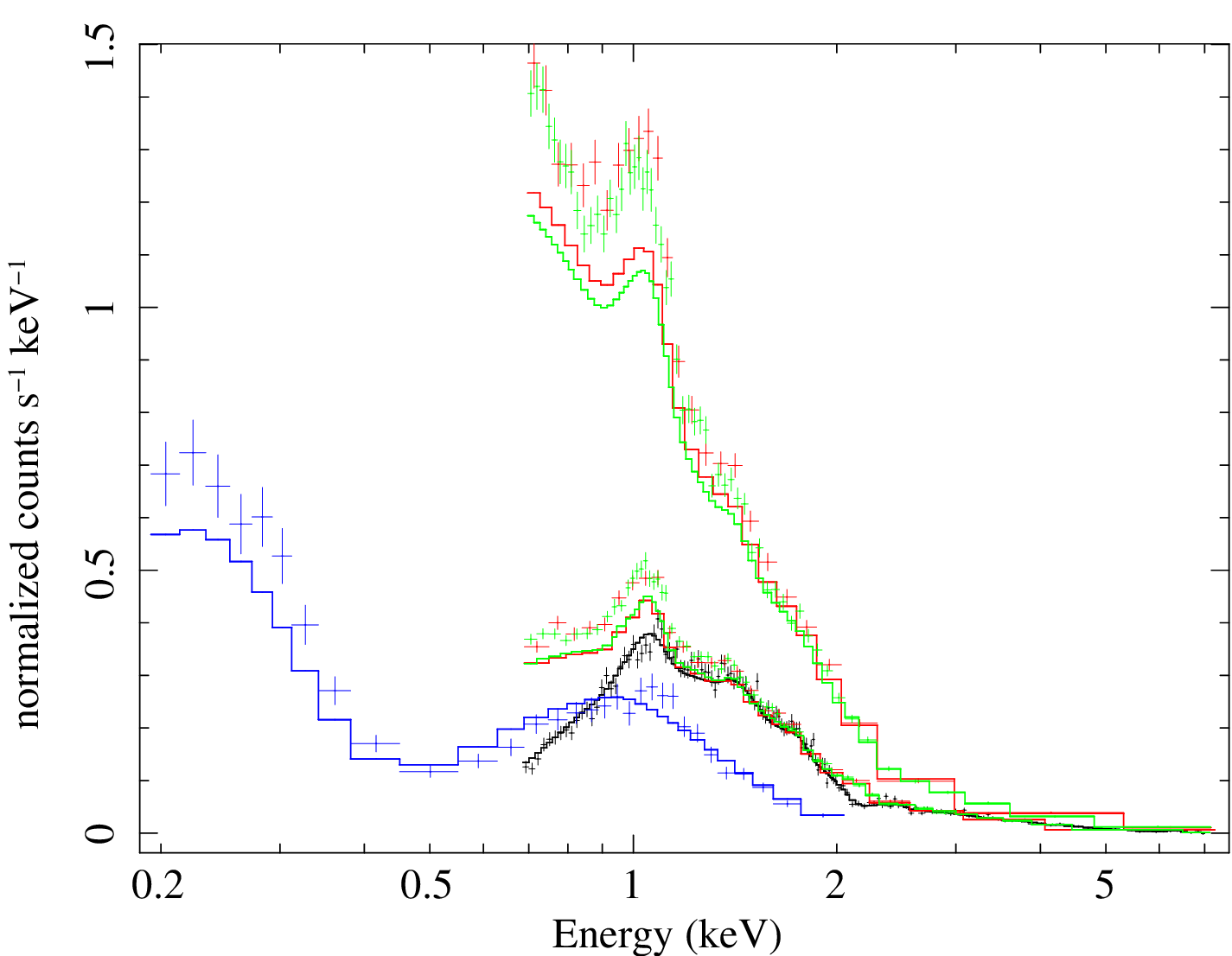}
\includegraphics[width=2.2in,angle=-90]{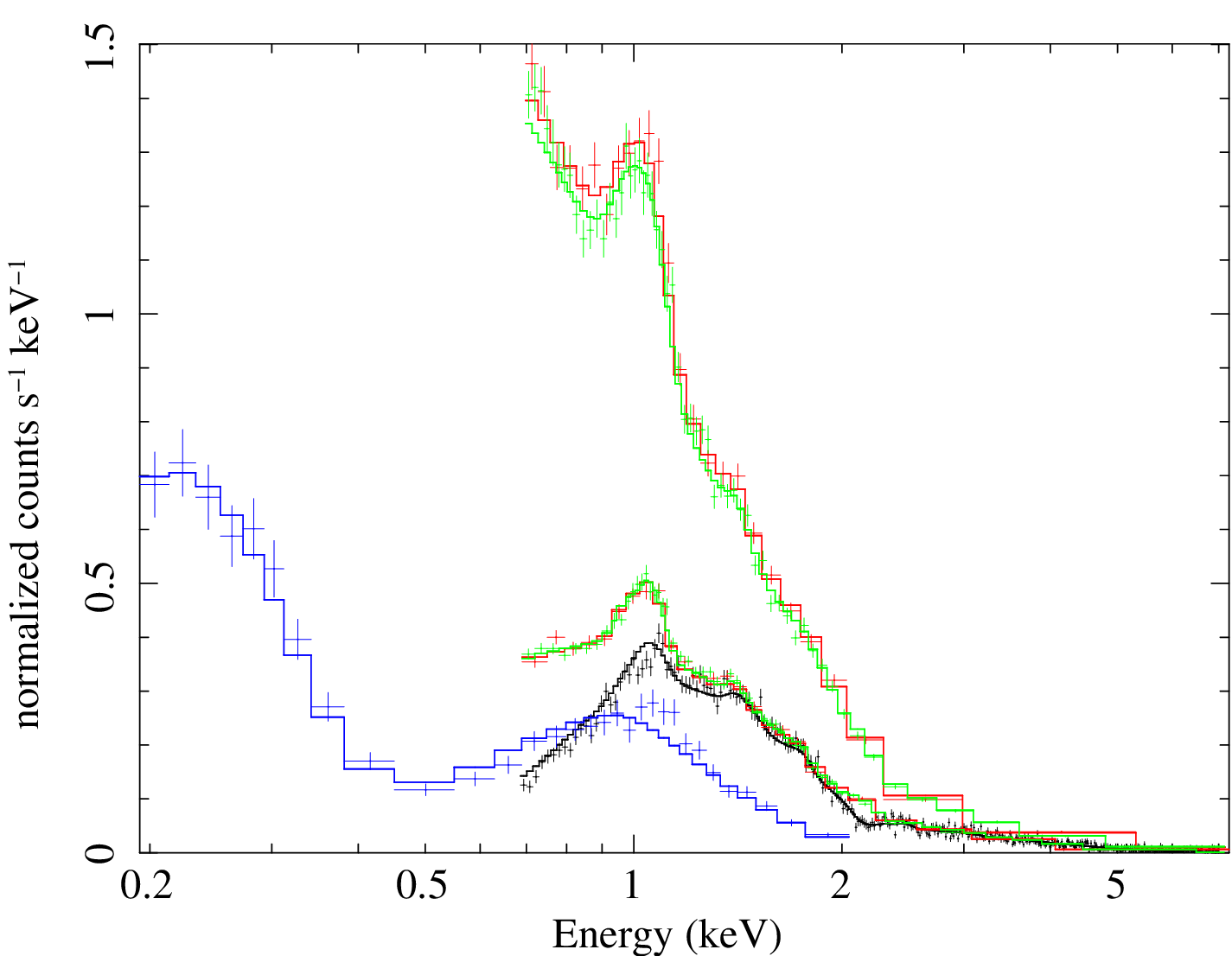}
\caption{Top: Spectra of the region 0-2 arcmin; Bottom: 
Spectra of the region 2-4 arcmin.
For both regions, \chandra\ spectra are in black, \xmm\ in green (observation 0147800101)` and red  (observation 0147800101),
MOS is the lower set of curves, pn is the higher set, because of higher effective area, 
and \rosat\ in blue\label{fig:spectra}.
On the Left: Best-fit models are for the 1-APEC fit to the hard band, and the extrapolation to the soft band
shows the excess emission above the hard band model.
On the right: Best-fit models are for the 2-APEC model for the hot gas, plus a warm APEC model of fixed $A=0$, as described
in Table~\ref{tab:data-2apec-soft}. Spectra were rebinned for clarity.}
\end{figure*}

We also fit the whole band spectra to the customary 2-APEC model representative of the hot gas, plus
a soft component to model the soft excess emission. The soft models are an APEC
plasma with zero abundance, an APEC plasma with $A=0.3$ abundance, and a power-law model.
For the hot gas, the normalization was left free among the three spectra; the normalization and
the temperature, or spectral index, of the soft
component were linked among the spectra. For all regions, the normalization of the
thermal component at half the temperature of the peak ICM model is found
to be consistent with zero.
The results are shown in Table~\ref{tab:data-2apec-soft}, and indicate that the addition
of the soft model leads to a significant improvement of the fit, when compared
to a fit to the same spectra without the soft model.
Based on the goodness of fit, the data do not indicate a strong preference for 
any of the three models for the soft excess.
We notice that the best-fit temperatures of the thermal model with $A=0.3$ abundance are significantly
lower than those of the model with primordial abundances. This is likely due to the fact that the 
soft component is truly `soft', i.e., mainly present in the \rosat\ 1/4 keV band, 
since emission from gas at $kT\geq 0.25$ ~keV
has a stronger line contribution in the higher \chandra\ and \xmm\ bandpass, 
apparently not present in the data.
Representative cases of the fit to the 0-2 and 2-4 arcmin spectra with the additional $A=0$ thermal component 
are shown in the right-hand panels of Figure~\ref{fig:spectra}. 
The fact that \chandra\ and \xmm\ do not agree in their 0.7-2~keV
fluxes results in a slight overprediction of the 0.7-1~keV \chandra\ flux, while the \xmm\ data in the same band
are fit accurately by this model. Other models to the soft excess (the $A=0.3$ thermal model and 
the non-thermal model)
give similar results, and are not shown in Figure~\ref{fig:spectra}.

We calculate the soft excess fluxes associated with the $A=0$ additional thermal component in the 0.3-1~keV band,
and find them to be respectively 1.4, 1.5 and 0.7 $\times 10^{-11}$ erg cm$^{-2}$ s$^{-1}$ deg$^{-2}$
for the three regions 0-2, 2-4 and 4-6 arcmin radius.
These fluxes are lower than those found by \cite{werner2007} based on \xmm\ and \suzaku\ data, consistent
with the fact that the new value of $N_H$ implies reduced soft excess emission.

A source of uncertainty associated with these model fits is the choice of  atomic physics
included in the APEC model \citep{smith2001}, which differ from other codes
\citep[e.g., the MEKAL/SPEX code, ][]{mewe1985} for certain emission lines such as
those of O~VII. We tested this effect by fitting the 2-4 armin region using the 
MEKAL code for the warm component, and obtain a fit in which the 
best-fit temperature is $kT \leq 0.11$~keV (the MEKAL model covers the temperature
range $kT\geq 0.081$~keV), with a normalization of $0.61\pm^{0.13}_{0.16}$
(in same units as those in Table~\ref{tab:data-2apec-soft}), for 
 $\chi^2_{min}=1466.5$ for the  
same number of degrees of freedom as the APEC model. The fit has therefore
the same statistical quality, and the higher best-fit temperature
implies a reduced emission integral for the hot gas, since a cooler
gas requires a higher emission integral to produce the same number
of R2 band counts. The implications of this
systematic uncertainty in the determination of the soft excess temperature
are  discussed in Section~\ref{sec:interpretation}.


\section{Interpretation of the soft X-ray fluxes}
\label{sec:interpretation}
Our results on the soft excess emission from \as1101\ show that the cluster
has a level of excess emission at the $\leq$20\% of the thermal virial gas,
in agreement with the typical soft excess emission from a large sample of
\rosat\ clusters analyzed in \cite{bonamente2002}.
In this Section we investigate two possible scenarios for the soft excess in Abell~S1101,
a thermal origin from diffuse filaments, and a non-thermal origin.

\subsection{Thermal origin from large-scale filaments}
\label{sec:filaments}
An  explanation for the excess emission in \as1101\ and for the other \rosat\
clusters analyzed in \cite{bonamente2002} that feature a 10-20\% soft excess, is that
the emission originates from large-scale warm filaments seen 
in projections against the hot cluster gas.
Filamentary structures of warm-hot intergalactic medium (WHIM)
 are routinely seen in numerical simulations
of large scale structure formation \citep[e.g.][]{dave2001}, and even tentatively
detected via their low surface brightness soft X-ray emission in
a region between two neighboring galaxy clusters \citep{werner2008}, and in 
the analysis of a sample of filaments of galaxies \citep{fraser2011}, although the
emission  reported in the latter is in a higher energy band (0.9-1.3~keV).
These filaments, are believed to contain a significant amount of the universe's baryons
\citep[e.g.][]{cen1998}, and have a temperature
($T \sim 10^5-10^7$ K) that makes their detection very challenging.
Attempts at their detection in absorption are likewise difficult
\citep{zappacosta2005, zappacosta2010,nicastro2005,nicastro2010,fang2010,buote2009,
takei2007,kaastra2006}, primarily because it requires a bright source
suitably located in their background, and the presence of metals.
The filamentary origin was also studied in \cite{bonamente2005} based on
\xmm\ data of \as1101, using the higher  $N_H$ value.

We use the thermal fits described in Section \ref{sec:2apec-soft} to estimate the 
length of the putative filaments responsible for the soft emission component.
We assume a simple geometry of cylindrical filaments
of constant density directed along the line of sight, with footprints equal to
the area of each annulus. In XSPEC, the emission integral of the putative warm gas is proportional
to the normalization constant of the spectrum via
\begin{equation}
N = \frac{10^{-14}}{4 \pi D_A^2 (1+z)^2} \int n_e n_H dV \;\;\text{cm$^{-5}$}
\end{equation} 
where $D_A$ is in cm, $n_e$ and $n_H$ are densities in cm$^{-3}$, $V=A \cdot L$
is the volume of the emitting gas, $A$ the area of the annulus, and $L$ the
length  along the sightline. We therefore use the normalization constant of the warm model
to calculate the length $L$ of the filaments, assuming a fiducial density of 
filament electrons of
$n_{fil}=10^{-4}$ cm$^{-3}$, and   $n_e/n_H$ = 1.2, appropriate for a highly ionized
plasma with low metal abundances.
The results, shown in Table~\ref{tab:length-pressure}, indicate that the filaments would be of
order few Mpc, perhaps somewhat longer for the central region. 
Given that our estimates assume a constant filament density, and that 
the estimated length depends on the density as $ L \propto N \propto n_{fil}^{-2}$, 
the length 
estimate for the central annulus would be reduced if the filaments are denser in that direction.
A recent measurement of filament density by \cite{fraser2011} finds that
\rosat\ observations in the direction of galaxy filaments is consistent with the emission
from sub-virial gas at a density of few times $10^{-4}$ cm$^{-3}$, for a plasma temperature of
1~keV. If similar densities apply to the WHIM model presented in this paper, which features
substantially lower temperatures, we calculate that
a filament  density $3 \times 10^{-4}$ cm$^{-3}$ would reduce the best estimates of the length towards
the 0-2 arcmin region to  a more plausible 3--5 Mpc. 

Our estimates based on this analysis of \chandra, \xmm\ and \rosat\ data
with the new $N_H$ measurement of \cite{kalberla2005} 
provide a scenario in which the soft X-ray emission from \as1101\ is in fact consistent
with an origin from relatively dense WHIM filaments. This re-analysis therefore brings \as1101\ in line
with the results obtained for a large sample of clusters observed by \rosat\ \citep{bonamente2002},
in which the excess emission is of the order 10-20\% of that from the hot ICM.
This thermal explanation was also found to be  plausible for the
large-scale halo of soft X-ray radiation around the brightest nearby
cluster, the Coma cluster \citep{bonamente2003,bonamente2009}.

Our \rosat\ soft X-ray data do not indicate strong preference between a thermal
model of the warm gas with primordial abundances, or one with presence of metals ($A=0.3$,
see Table~\ref{tab:data-2apec-soft}). We calculate limits to the
presence of O~VII emission lines in the \rosat\ data following the \cite{werner2007} method
by placing a narrow Gaussian emission line (of width parameter $\sigma=0.1$~keV) 
at the expected line energy, above the non-thermal model described in Section~\ref{sec:2apec-soft}.
We calculate 
2$\sigma$ upper limits for the three regions of Table~\ref{tab:data-2apec-soft} as
respectively 1.2, 1.4 and 1.0 $\times 10^{-6}$ photons cm$^{-2}$ s$^{-2}$ arcmin$^{-2}$, which
are similar to the limits found by \cite{werner2007}. 

A significant source of systematic error in the estimate of filament lengths is 
associated with the determination of the filament temperature. In fact, even the
\rosat\ soft X-ray band is not ideally suited for the measurement of temperatures
below approximately a few times 100~eV.
The intensity of emission lines  from a plasma at sub-virial temperatures is also a source
of uncertainty when measuring the temperature of  warm gas.  
As an example of this source of uncertainty, in Section~\ref{sec:2apec-soft} we
have shown that the 2-4 arcmin annulus excess, when fit to a MEKAL model
instead of an APEC model,
yields a higher filament temperature and an emission integral
that is lower by one order of magnitude; this implies filament lengths that are lower
by the same amount. Because of the difficulties in measuring the filament temperature,
the filament lengths provided in Table~\ref{tab:length-pressure}
should be regarded only as order-of-magnitude estimates, e.g., these estimates could be significantly lower
if the gas is at a somewhat higher temperature.
 Only a detection of specific emission lines from such
ions as O~VII will provide an accurate estimate of the temperature
of the gas associated with the soft excess component.

The numerical simulations by \cite{cheng2005} find that the soft excess emission
may be explained with the presence of low-entropy, high-density gas, rather than
the diffuse, high-entropy gas assumed in a WHIM filament model. Our observations
do not have the resolution to probe whether the soft excess emitter is truly diffuse
or localized in dense regions. If the \cite{cheng2005} model is correct, the
mass of warm gas required to explain the excess would be lower than that implied by the
filament model.

\subsection{Non-thermal origin of the soft excess}

Inverse Compton scattering
with a population of relativistic
electrons can shift CMB photons to X-ray energies,
and give rise to the soft excess emission \citep[e.g.][]{sarazin1998,bonamente2005,lieu2010}.
A power-law distribution of cosmic ray electrons with differential
number index $\mu$ generates a power-law  spectrum with
differential photon number index $\alpha=(\mu+1)/2$. Therefore, 
the parameters of the power-law model described in Section~\ref{sec:2apec-soft}
can be used to infer the energetics of relativistic electrons that may
give rise to the soft excess.

In Table~\ref{tab:length-pressure} we calculate the  non-thermal
pressure of the emitter, using the results of Table~\ref{tab:data-2apec-soft}, and
Equation~4 of \cite{bonamente2005}. 
For these calculations we use the measured non-thermal unabsorbed luminosities of respectively
3.6, 2.9 and 2.2 $\times 10^{42}$ erg s$^{-1}$ for the three regions, in the \rosat\ 0.2-2 keV band, 
and symmetric errors for the fit parameters. These estimates for the ratio of non-thermal
to thermal
pressure are lower than those based on the higher
value of $N_H$ \citep{bonamente2005,werner2007}, consistent with the reduction
in soft excess fluxes because of the lower $N_H$ value from the LAB survey.
The result is that the putative non-thermal electrons would feature
a pressure that is less than 1\% of the thermal pressure, and therefore provide a marginal
contribution to the overall cluster energetics.

\begin{table*}[!ht]
\footnotesize
\centering
\begin{tabular}{lcccccc}
\hline 
Annulus & \multicolumn{2}{c}{Hot gas model} & \multicolumn{4}{c}{R2 band fluxes (c/s)} \\
(arcmin) &  kT (keV) & A  & detected flux  & model flux & $\eta$ &  \% above background \\
         &           &    & \multicolumn{2}{c}{$10^{-2}$ c/s} & \\
\hline
0-2  & 2.70 $\pm^{0.02}_{0.02}$ & 0.38 $\pm^{0.01}_{0.01}$ & 19.6 $\pm 0.4$ & 17.0 $\pm 0.2$ & 0.16$\pm^{0.05}_{0.03}$  & 98.0 \\ 
2-4  & 2.80 $\pm^{0.05}_{0.05}$ & 0.26 $\pm^{0.02}_{0.02}$ & 6.7 $\pm 0.3$ & 5.9  $\pm 0.2$  & 0.14$\pm0.06$  & 82.9 \\ 
4-6  & 2.63 $\pm^{0.10}_{0.10}$ & 0.32 $\pm^{0.07}_{0.06}$ & 3.5 $\pm 0.2$ & 3.1  $\pm 0.3$  & 0.12$\pm0.12$  & 59.4 \\ 
6-8  & 1.63 $\pm^{0.16}_{0.15}$ & 0.03 $\pm^{0.09}_{0.03}$ & 2.4 $\pm 0.2$ & 2.6  $\pm 0.6$  & -0.07$\pm0.30$  & 40.1 \\ 
8-10 &1.93 $\pm^{0.58}_{0.35}$ & 0.00 $\pm^{0.24}_{0.00}$ & 1.4 $\pm 0.2$ & 1.5  $\pm 0.5$   & -0.05$\pm0.50$  & 22.6 \\ 
\hline
\hline
\end{tabular}
\caption{Best-fit \chandra, \xmm\ and \rosat\ hot gas parameters in hard band (2-7 keV for \chandra\ and \xmm,
1-2 keV for \rosat) and R2 band fluxes.\label{tab:data}}
\end{table*}

\begin{table*}[!ht]
\footnotesize
\centering
\begin{tabular}{lcccccc}
\hline
Annulus & \multicolumn{3}{c}{Hot gas model} & \multicolumn{3}{c}{R2 band fluxes (c/s)} \\
(arcmin) &  kT (keV) & {1/2-$kT$ fraction} & A  & detected flux  & model flux & $\eta$ \\
         &           &    & & \multicolumn{2}{c}{$10^{-2}$ c/s} \\
\hline
0-2 & 2.88 $\pm^{0.07}_{0.05}$ & 0.37 $\pm^{0.01}_{0.01}$ & 0.24 $\pm 0.06$ & 19.6 $\pm 0.4$ & 16.7 $\pm 0.2$ & 0.17$\pm{0.03}$ \\
2-4 & 3.28 $\pm^{0.18}_{0.28}$ & 0.25 $\pm^{0.02}_{0.02}$ & 0.84 $\pm 0.28$ & 6.7 $\pm 0.3$ & 6.2 $\pm 0.3$   & 0.08$\pm^{0.05}_{0.06}$ \\
4-6 & 2.17 $\pm^{0.19}_{0.39}$ & 0.33 $\pm^{0.07}_{0.07}$ & 0.29 $\pm 0.16$ & 3.5 $\pm 0.2$  & 3.1 $\pm 0.3$  & 0.14$\pm^{0.13}_{0.12}$ \\
\hline
\hline
\end{tabular}
\caption{Best-fit \chandra\ + \xmm\ hot gas parameters and R2 band fluxes, using the
2-APEC model with second thermal component fixed at 1/2 the temperature of the hot component.\label{tab:data-2apec}}
\end{table*}

\begin{table*}[!ht]
\footnotesize
\centering
\begin{tabular}{lccccc}
\hline
Annulus & \multicolumn{2}{c}{Hot gas model} & \multicolumn{3}{c}{R2 band fluxes (c/s)} \\
(arcmin) &  kT (keV) & A  & detected flux  & model flux & $\eta$ \\
         &           &    & \multicolumn{2}{c}{$10^{-2}$ c/s} & \\
\hline
0-2  & 2.80 $\pm^{0.01}_{0.02}$ & 0.47 $\pm^{0.01}_{0.01}$ & 19.6 $\pm 0.4$ & 17.7 $\pm 0.2$ & 0.11$\pm^{0.02}_{0.03}$\\
2-4  & 2.86 $\pm^{0.04}_{0.04}$ & 0.32 $\pm^{0.03}_{0.03}$ & 6.7 $\pm 0.3$ & 6.1 $\pm 0.3$   & 0.10$\pm^{0.06}_{0.05}$\\
4-6  & 2.69 $\pm^{0.10}_{0.08}$ & 0.37 $\pm^{0.08}_{0.07}$ & 3.5 $\pm 0.2$ & 3.2 $\pm 0.3$   & 0.09$\pm{0.12}$ \\
6-8  & 1.53 $\pm^{0.21}_{0.10}$ & 0.09 $\pm^{0.11}_{0.05}$ & 2.4 $\pm 0.2$ & 2.5 $\pm 0.6$   & -0.02$\pm0.30$  \\
8-10 & 1.79 $\pm^{0.53}_{0.28}$ & 0.13 $\pm^{0.29}_{0.13}$ & 1.4 $\pm 0.2$ & 1.4 $\pm 0.7$   & -0.04$\pm0.50$  \\
\hline
\hline
\end{tabular}
\caption{Best-fit \chandra\ + \xmm\ hot gas parameters and R2 band fluxes,
using the \cite{balucinska1992} cross-sections for Galactic absorbing material .\label{tab:data-phabs}}
\end{table*}

\begin{table*}[!ht]
\footnotesize
\centering
\begin{tabular}{lccccccccc}
\hline
Annulus &  kT (keV)  &  A & detected flux  & model flux & $\eta$ & $\chi^2$ (dof) \\
(arcmin) &           &    &  \multicolumn{2}{c}{$10^{-2}$ c/s} & &  \\
\hline
\multicolumn{6}{c}{\xmm\ \text{plus} \rosat}\\
0-2 & $2.55\pm^{0.01}_{0.01}$ & $ 0.43\pm^{0.00}_{0.00}$ & $19.64\pm 0.41$ & $16.32 \pm 0.28$ & 0.20$\pm$0.03  & $2050.84$ (1882) \\
2-4 & $2.56\pm^{0.03}_{0.03}$ & $ 0.25\pm^{0.01}_{0.01}$ & $6.70\pm 0.26$ & $5.57 \pm 0.30$   & 0.20$\pm^{0.08}_{0.07}$ & $1009.08$ (1106)  \\
4-6 & $2.37\pm^{0.07}_{0.06}$ & $ 0.16\pm^{0.01}_{0.01}$ & $3.52\pm 0.22$ & $3.15 \pm 0.50$   & 0.12$\pm$0.17  & $678.93$ (748)  \\
\hline
\multicolumn{6}{c}{\chandra\ \text{plus} \rosat}\\
0-2 & $2.69\pm^{0.01}_{0.01}$ & $ 0.46\pm^{0.01}_{0.01}$ & $19.64\pm 0.41$ & $16.01 \pm 0.35$ & 0.23$\pm^{0.03}_{0.04}$ & $763.34$ (542) \\ 
2-4 &  $2.74\pm^{0.06}_{0.03}$ & $ 0.22\pm^{0.02}_{0.02}$ & $6.70\pm 0.26$ & $5.61 \pm 0.33$  & 0.20$\pm^{0.07}_{0.08}$ & $513.56$ (497) \\ 
4-6 &  $3.30\pm^{0.96}_{0.63}$ & $ 0.11\pm^{0.03}_{0.03}$ & $3.52\pm 0.22$ & $3.04 \pm 1.29$  & 0.16$\pm0.43$ & $409.64$ (464) \\ 
\hline
\hline
\end{tabular}
\caption{Best-fit hot gas parameters and R2 band fluxes, 
from a whole band fit (\xmm, \chandra: 0.7-7 keV; \pspc: 0.5-2~keV).\label{tab:data-chandra-xmm-rosat-whole}}
\end{table*}

\begin{table*}[!ht]
\footnotesize
\centering
\begin{tabular}{lccccccc}
\hline
Annulus & \multicolumn{3}{c}{Hot gas model ($N_H=1.15\times10^{20}$ cm$^{-2}$)} & \multicolumn{4}{c}{Free-$N_H$ model} \\
(arcmin) &  kT  &  A & $\chi^2$ (dof)   & kT & A &  $N_H$  &  $\chi^2$ (d.o.f) \\
         & (keV)   &    & & (keV)  & &  ($10^{20}$ cm$^{-2}$)   & \\
\hline
0-2 & 2.60 $\pm^{0.01}_{0.01}$ & 0.43 $\pm^{0.01}_{0.01}$ & 2973.8 (2331) & 
	2.64$\pm^{0.01}_{0.01}$ & 0.44 $\pm^{0.01}_{0.01}$ & 0.85 $\pm$ 0.04 & 2837.0 (2330)    \\
2-4 & 2.70 $\pm^{0.04}_{0.03}$ & 0.24 $\pm^{0.01}_{0.01}$ & 1542.8 (1560) &
 	2.72$\pm^{0.02}_{0.02}$ & 0.25 $\pm^{0.01}_{0.01}$ & 0.68 $\pm$ 0.07 & 1503.3 (1559)\\
4-6 & 2.38 $\pm^{0.06}_{0.07}$ & 0.15 $\pm^{0.01}_{0.01}$ & 1103.3 (1216) &
  2.39  $\pm^{0.03}_{0.03}$ & 0.15 $\pm^{0.02}_{0.01}$ & 0.84$\pm^{0.11}_{0.16}$ & 1097.3 (1215) \\
\hline
\hline
\end{tabular}
\caption{Best-fit $N_H$ value obtained by a fit to the 
\chandra, \xmm\ and \pspc\ whole-band spectra, using the
2-APEC model with second thermal component fixed at 1/2 of the temperature of hot component.\label{tab:data-2apec-freeNH}}
\end{table*}

\begin{table*}[!ht]
\footnotesize
\centering
\begin{tabular}{lcccccc|c}
\hline
Annulus & \multicolumn{3}{c}{Hot gas model}  & \multicolumn{2}{c}{Soft component} & &  Hot gas only \\
(arcmin) &  kT (keV)  &  A &  Normalization  &     &   & $\chi^2$ (dof) & $\chi^2$ (dof) \\
\hline
        & \multicolumn{3}{c}{} &  \multicolumn{2}{c}{APEC model with $A=0.0$} & \\
        & \multicolumn{3}{c}{} &  kT     &  Normalization &   \\
0-2 & 2.62 $\pm^{0.01}_{0.01}$ & 0.43 $\pm^{0.00}_{0.00}$ & 19.14 $\pm 0.06$ & 0.10 $\pm^{0.02}_{0.03}$ & 11.77 $\pm^{10.03}_{2.79}$ &  2827.3 (2329) & 2896.7 (2331)\\ 
2-4 & 2.73 $\pm^{0.02}_{0.02}$ & 0.26 $\pm^{0.01}_{0.01}$ & 4.35 $\pm 0.04$ & 0.23 $\pm^{0.04}_{0.04}$ & 1.85 $\pm^{0.53}_{0.40}$ &  1457.5 (1485) & 1502.1 (1487)\\ 
4-6 & 3.11 $\pm^{0.19}_{0.30}$ & 0.28 $\pm^{0.08}_{0.04}$ & 0.84 $\pm 0.17$ & 1.00 $\pm^{0.32}_{0.37}$ & 0.80 $\pm^{0.30}_{0.19}$ &  1070.8 (1203) & 1094.1 (1205)\\ 
\hline
        & \multicolumn{3}{c}{} &  \multicolumn{2}{c}{APEC model with $A=0.3$} & \\
	& \multicolumn{3}{c}{} &  kT     &  Normalization &   \\
0-2 & 2.62 $\pm^{0.01}_{0.01}$ & 0.43 $\pm^{0.00}_{0.00}$ & 19.15 $\pm 0.06$ & 0.06 $\pm^{0.01}_{0.01}$ & 7.54 $\pm^{7.65}_{2.84}$ &  2822.0 (2329) & 2896.7 (2331)\\ 
2-4 & 2.69 $\pm^{0.04}_{0.03}$ & 0.24 $\pm^{0.01}_{0.01}$ & 4.44 $\pm 0.03$ & 0.05 $\pm^{0.03}_{0.01}$ & 3.58 $\pm^{7.04}_{2.15}$ &  1466.7 (1485) & 1502.1 (1487)\\ 
4-6 & 2.44 $\pm^{0.07}_{0.06}$ & 0.18 $\pm^{0.02}_{0.02}$ & 1.84 $\pm 0.06$ & 0.23 $\pm^{0.07}_{0.04}$ & 0.12 $\pm^{0.12}_{0.06}$ &  1082.7 (1203) & 1094.1 (1205)\\ 
\hline
        & \multicolumn{3}{c}{} &  \multicolumn{2}{c}{Power-law model} & \\
	& \multicolumn{3}{c}{} &  $\alpha$      &  Normalization &   \\
0-2 & 2.63 $\pm^{0.01}_{0.01}$ & 0.44 $\pm^{0.01}_{0.01}$ & 19.01 $\pm 0.12$ & 3.98 $\pm^{0.79}_{0.61}$ & 0.02 $\pm^{0.03}_{0.01}$ &  2830.2 (2329) & 2896.7 (2331) \\ 
2-4 & 2.74 $\pm^{0.03}_{0.02}$ & 0.28 $\pm^{0.01}_{0.01}$ & 4.16 $\pm 0.11$ & 2.72 $\pm^{0.29}_{0.23}$ & 0.06 $\pm^{0.03}_{0.02}$ &  1449.0 (1485) & 1502.1 (1487)\\ 
4-6 & 2.33 $\pm^{0.13}_{0.14}$ & 0.19 $\pm^{0.02}_{0.02}$ & 1.62 $\pm 0.09$ & 2.17 $\pm^{0.23}_{0.16}$ & 0.07 $\pm^{0.02}_{0.02}$ &  1078.2 (1203) & 1094.1 (1205)\\ 
\hline
\end{tabular}
\caption{
Fit to a 2-APEC model with second thermal component fixed at 1/2 the temperature  of the hot component,
plus an additional soft component (APEC, or a power-law model). 
Fit is to \chandra, \xmm\ and \pspc\ whole-band spectra.
The normalization of the warm thermal component is in units of $10^{-11}/(D_A^2(1+z)^2) \int n_e n_H dV$,
and the normalization of the power-law model is in units of $10^{+3}$~photons keV$^{-1}$ cm$^{-2}$ s$^{-1}$ at 1 keV.
}
\label{tab:data-2apec-soft}
\end{table*}

\begin{table*}[!ht]
\footnotesize
\centering
\begin{tabular}{lcccccc}
\hline
Annulus & \multicolumn{2}{c}{Filament length (Mpc)} & \multicolumn{2}{c}{Average pressure ($10^{-11}$ erg cm$^{-3}$)} \\
(arcmin)   & $A=0$ & $A=0.3$ & Hot ICM & Non-thermal component \\ 
\hline
  0-2 & 48.00 $\pm^{40.93}_{11.39}$ & 30.76 $\pm^{31.19}_{11.58}$ & 26.0$\pm$0.11 &$0.06\pm0.06$  \\
  2-4 & 2.52 $\pm^{0.73}_{0.55}$ & 4.86 $\pm^{9.58}_{2.92}$ & $4.78\pm 0.05$ & $0.006\pm 0.002$ \\
  4-6 & 0.66 $\pm^{0.25}_{0.16}$ & 0.10 $\pm^{0.10}_{0.05}$ & $1.54\pm 0.08$ & $0.0015\pm 0.0005$ \\
\hline
\end{tabular}
\caption{Estimates of filament length and thermal-to-non thermal pressure
for the models of Table~\ref{tab:data-2apec-soft}. Filament lengths assume $n=10^{-4}$ cm$^{-3}$.}
\label{tab:length-pressure}
\end{table*}



\section{Conclusions}
\label{sec:conclusions}
In this paper we have presented spatially resolved X-ray spectroscopy of \as1101\ with
\chandra, \xmm\ and \rosat. Our observations show that calibration uncertainties
in the 0.7-2 keV X-ray band remain between the \chandra\ and \xmm\ detectors, with the
\xmm\ whole-band best-fit temperatures typically lower than those measured by \chandra,
as also found by \cite{nevalainen2010}.

Our analysis confirms the presence of
soft excess emission in the \rosat\ 1/4 keV band, when the lower $N_H$ value of \cite{kalberla2005} is used,
and regardless of calibration uncertainties between \chandra\ and \xmm\ in the 0.7-2 keV band.
The  revised soft excess
fluxes that we derived show that the level of soft X-ray excess emission in all clusters
studied to date does not typically exceed $\sim$20\% of the hot gas emission
in regions that are within the virial radius.
In fact, using the scaling relation
$r_{500}=(0.796\pm0.015)/(h E(z)) (T/5 \text{keV}))^{(1.61\pm0.11)/3}$~Mpc
from \cite{vikhlinin2006}, a $kT=3$~keV cluster like \as1101\ has
$r_{500} \simeq$ 1~Mpc, and the \rosat\
1/4 keV band excess is certainly within 6~arcmin, or approximately $\sim$400~kpc;
a similar situation applies
to other clusters studied with \rosat\ in \cite{bonamente2002}.
A possible exception is the Coma cluster, for which we have shown that there is excess emission
out to $\sim$5~Mpc \citep{bonamente2009}, and for a cluster at $kT=8$~keV such as Coma,
$r_{500} \simeq$ 1.9~Mpc. The case of Coma may be exceptional in that it is one of the closest and brightest clusters,
featuring  higher S/N observations  than any of the other clusters, and therefore it
is possible that higher resolution soft X-ray observations near the virial radius may reveal that other clusters have soft X-ray halos
that exceed the extent of the virialized gas.

The soft excess emission was interpreted as both thermal and non-thermal emission, and
found that our data are consistent with both models for the excess.
A possible origin for the soft excess is
from WHIM filaments that are expected to converge towards massive galaxy clusters,
and therefore become visible in X-rays even in the absence of metal lines.
The best-fit temperature measured in this study
are typically lower than the WHIM probed in other X-ray studies \citep[e.g.,][]{werner2008}.
In the case of a non-thermal origin, we find that the relativistic electrons responsible for the emission
would feature a pressure of less than 1\% of the thermal electrons.
The upcoming launch of eRosita \citep{predehl2010},
which improves on \rosat's large effective area at 1/4 keV and field of view, will enable
further tests of the presence of warm gas near clusters such as the ones
we have performed for this paper with \rosat.

\section*{Acknowledgments}
The authors thank the referee for a thorough review of the manuscript and many suggestions that
led to significant improvements to this paper.


\end{document}